\documentclass[varenna]{cimento}
\newcommand {\be} {\begin{eqnarray*}}
\newcommand {\ee} {\end{eqnarray*}}
\newcommand {\bea} {\begin{eqnarray}}
\newcommand {\eea} {\end{eqnarray}}
\newcommand{\ket}[1]{| {#1} \rangle}
\newcommand{\bra}[1]{\langle {#1} |}
\def\kb#1#2{|#1\rangle\!\langle #2 |}
\def\bk#1#2{\langle #1 | #2 \rangle}
\def\kb#1#2{|#1\rangle\!\langle #2 |}
\def\bk#1#2{\langle #1 | #2 \rangle}
\newcommand{\dg}{^\dagger}

\def\id{{\mathchoice {\rm 1\mskip-4mu l} {\rm 1\mskip-4mu l} {\rm
1\mskip-4.5mu l} {\rm 1\mskip-5mu l}}}

\usepackage[]{graphicx}
\usepackage{amsmath}
\usepackage{amsfonts}
\usepackage{amssymb}%
\usepackage{color}
\usepackage{hyperref}
\usepackage{bm}
\usepackage{graphicx}
\usepackage{amsbsy}

\title{Linear Optics Quantum Computation: an Overview}

\author{C.~R. Myers}

\institute{Institute for Quantum Computing and Physics Department,
  University of Waterloo, ON, N2L 3G1, Canada}

\author{R. Laflamme}

\institute{Institute for Quantum Computing and Physics Department,
  University of Waterloo, ON, N2L 3G1, Canada\\ Perimeter Institute
  for Theoretical Physics, 31 Caroline St N., Waterloo, ON, N2L 2Y5,
  Canada}

\begin{document}

\maketitle

\tableofcontents


\begin{figure}[h!]
\begin{center}
\includegraphics[width=1.0\textwidth]{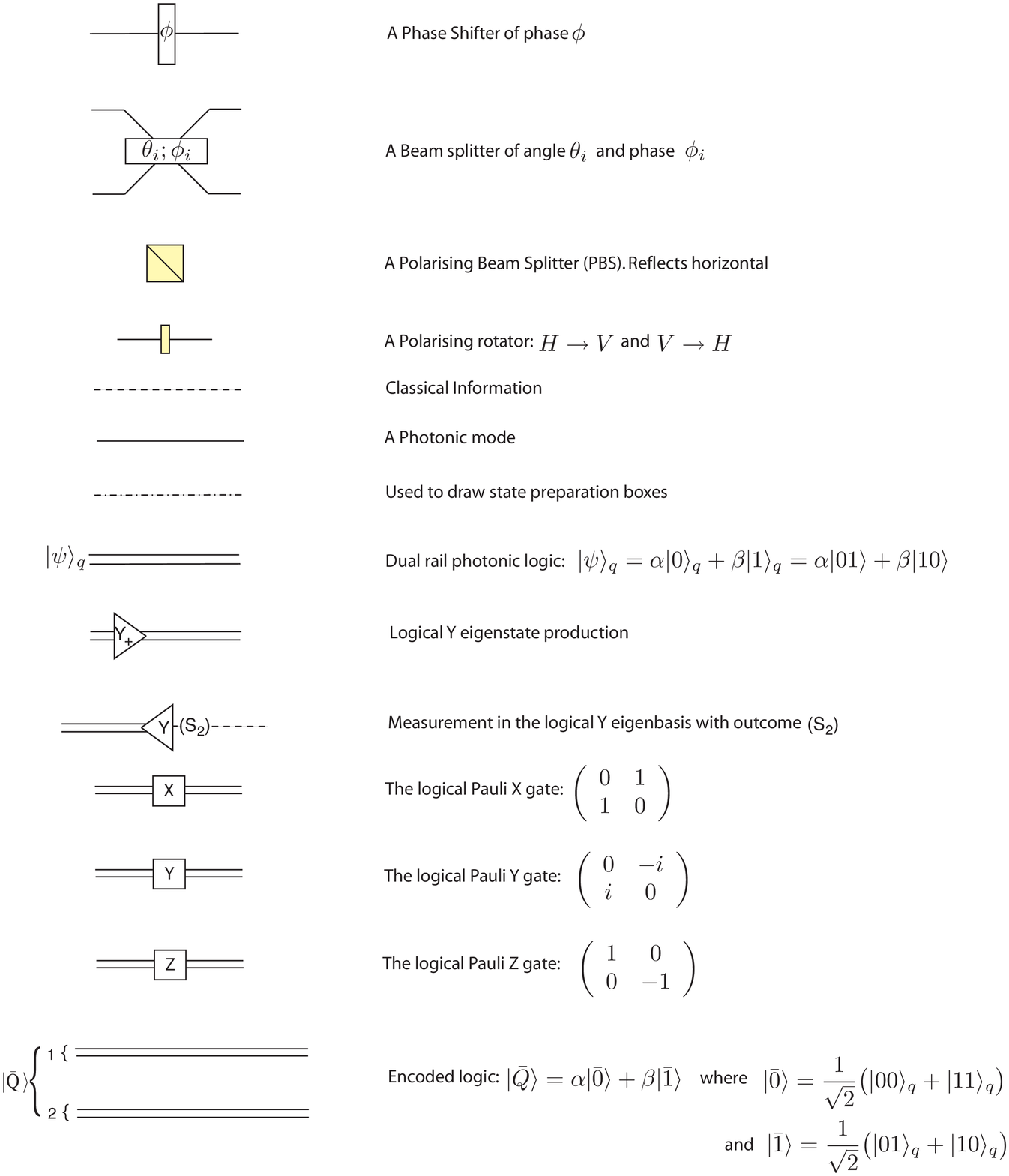}

\end{center}
\caption{\footnotesize Notation for the lectures. }
\label{NotationPic}
\end{figure}
\newpage
\section{Quantum Information Processing with linear optics}

There are at least three  ways to understand these lectures on linear optics 
quantum computing:
\begin{itemize}
\item these are engineering lectures on how to build an optical quantum computer.  We will give a step by step approach on how  to use beam splitters, phase shifters, single photon sources and detectors to efficiently simulate an ideal quantum computer. In this sense, these lectures can really be thought of as being an engineering recipe,
\item they can be thought of as lectures on quantum error correction. In order to use the simple elements mentioned above to build quantum information processing devices we will have to rely heavily on ideas and concepts of quantum error correction. By putting elements together and using detectors, a simple error model will emerge (projection of a qubit in the $Z$ basis) and we will use fault tolerant error correction to ensure that the quantum information is protected.  We will even calculate a value for the accuracy threshold,
\item or these can be seen as lectures on what quantum information is. These lecture will teach you that maybe we do not understand what quantum information really is. Some years ago it was thought that a quantum computation would be implemented by doing very precise unitary transformations. Here we will do very non-unitary operations (projections) on physical qubits but we will still be able to show that at a higher level this can mimic a unitary evolution efficiently.  Linear optics will allow us to do a set of transformation that can be simulated efficiently by a classical computer. It is only when we condition our operations on the results of measurements that the system becomes hard to be simulated classically.
\end{itemize}

\subsection{Quantum optics and quantum information}

These lectures will be divided into three sections. First an introduction to quantum optics which will include some historical notes on the use of quantum optics for quantum information processing (QIP).  The second section will focus on using linear optics with single photon sources and detectors (sometimes called the LOQC or KLM theory). The third section will focus on quantum error correction and LOQC.

\subsection{Quantum Computation}\label{sect:qunatumcomputation}

Quantum information promises to solve some problems that have no known efficient classical algorithms. This is done by harnessing the size of Hilbert space and the ability to manipulate the quantum state.

David DiVincenzo \cite{DiVincenzo00} has suggested a set of five
criteria that are sufficient for a device to be used as a quantum
computer:

\begin{enumerate}
\item A scalable physical system with well characterised qubits.
\item The ability to initialise the state of the qubits to a simple fiducial state, such as $\ket{000\ldots}$. 
\item A ``universal'' set of quantum gates such as generic one qubit gates and
a two qubit gate,
\item A qubit-specific measurement capability.
\item Long relevant decoherence times, much longer than the gate operation time. 
\end{enumerate}

\subsection{Why optics?}

Many physical systems have be proposed for quantum computation. Each
has its own advantages and disadvantages. For example, in an ion trap,
two qubit gates are relatively easy to implement. However, isolating
the ions from the environment is difficult, the ions' motion being vulnerable
to decoherence. In NMR quantum computing, the most successful proposal
to date, nuclear spin states can be handled directly. However, the
ability to initialise a simple fiducial state is difficult. Also, many
nuclear spins must be present in order for a sizable signal to be
observed.

The first proposal for quantum computing involved single photons as
qubits \cite{Milburn89}. One advantage for using single photons in QIP
is that quantum optics is a well developed field. Photons decohere
slowly, an ideal trait for quantum computing. Photons also travel
well, one reason why they are so widely used for
communication. Another advantage is that photons can be experimented
with at room temperature.

One of the major obstacles for using single photons for QIP is that photons do not interact directly, making two qubit gates very difficult. Mode matching with single photon is also a problem. Another
difficulty with using single photons is how to produce and detect them. To date the best single photon sources and detectors both have efficiencies well below any known threshold for scalable quantum computation.

\subsection{Quantum Optics}

\subsubsection{Classical Electromagnetic Field}

When we deal with electromagnetic waves classically we look for
solutions to the source free Maxwell eqs.
\begin{eqnarray}
  \nabla^{2} {\bf E}({\bf r},t) -\frac{1}{c} 
\frac{\partial^{2}}{\partial t^{2}} {\bf E}({\bf r},t)=0
 \end{eqnarray}
The solution have a plus and minus frequency part \cite{Walls, Scully} :  
\begin{eqnarray}
{\bf E}({\bf r},t)=i \sum_{k} 
\biggl( \frac{\hbar \omega_{k}}{2}\biggr)^{\frac{1}{2}}
   [a_{k} {\bf u}_{k}({\bf r})e^{-i \omega_{k} t} 
  + a^{*}_{k} {\bf u}^{*}_{k}({\bf r})e^{i \omega_{k} t}]
\end{eqnarray}
The ${\bf u}_{k}({\bf r})$ are orthogonal mode functions, usually
plane wave mode functions (
${\bf u}_{k}({\bf r})=\frac{\hat{{\bf e}}^{(\lambda)}}
{\sqrt{V}}\exp(i {\bf k}\cdot {\bf r})$, where
$\hat{{\bf e}}^{(\lambda)}$ is the unit polarisation vector and $V$ is
the volume), and the $a_{k}$ are dimensionless (complex) amplitudes.

The energy of a classical electromagnetic field is given by
\begin{eqnarray}
H=\frac{1}{2}\int_{V} ({\bf E}^{2}+{\bf B}^{2})d{\bf r} = 
  \sum_{k}\biggl(\frac{\hbar\omega_{k}}{2}\biggr) a_{k}  a^{*}_{k}
\end{eqnarray}

\subsubsection{Quantise}

We quantise the electromagnetic field by turning the coefficient
$a_{k}$ into operators and imposing the commutation relations:   
\begin{eqnarray}
&&[\hat{a}_{i},\hat{a}^{\dagger}_{j}]=\delta_{i,j} \\
&&[\hat{a}_{i},\hat{a}_{j}]=[\hat{a}^{\dagger}_{i},\hat{a}^{\dagger}_{j}]=0.
\end{eqnarray}
In  terms of these operators the energy is given by 
\begin{eqnarray}
\hat{H}=\sum_{k} \hbar \omega_{k} \bigl( \hat{a}^{\dagger}_{k}\hat{a}_{k} +\frac{1}{2}\bigr).
\end{eqnarray}
If we let 
\begin{eqnarray}
\hat{a}=\frac{1}{\sqrt{2}}\bigl( \sqrt{\frac{m\omega}{\hbar}}\hat{x}+i\frac{1}{\sqrt{m\hbar\omega}}\hat{p}\bigr)\label{aasxpEqn}
\end{eqnarray}
the energy of the electromagnetic field becomes 
\begin{eqnarray*}
\hat{H}=\frac{\hat{p}^{2}}{2m}+\frac{1}{2}m\omega^{2}\hat{x}^{2}
\end{eqnarray*}
from which we can recognise the harmonic oscillator.

We call $\hat{a}$ an annihilation operator and $\hat{a}^{\dagger}$ a
creation operator, the reason for this will be shown in eqs.
(\ref{CreationEqn}) and (\ref{AnnihilationEqn}).

The eigenstates of $\hat{H}$ are labeled $\ket{n}$ and we call them
Fock states. We define the number operator $\hat{n}$ by
$\hat{n}\ket{n}=\hat{a}^{\dagger}\hat{a}\ket{n}=n\ket{n}$.

What is the effect of $\hat{a}$ on $\ket{n}$? We can answer this by
finding the number of photons in $\hat{a}\ket{n}$:
\begin{eqnarray*}
\hat{n}\bigl(\hat{a}\ket{n}\bigr)=\hat{a}^{\dagger}\hat{a}^{2}\ket{n}
=\bigl( \hat{a}\hat{a}^{\dagger}\hat{a}-\hat{a}\bigr)\ket{n}
=\bigl(n-1\bigr) \hat{a}\ket{n}
\end{eqnarray*}
So $\hat{a}\ket{n}$ is a Fock state with $n-1$ photons. We define this
to be $A\ket{n-1}$. In a similar way we can show that
$\hat{a}^{\dagger}\ket{n}$ is a Fock state with $n+1$ photons,
$B\ket{n+1}$. But what is the form of $A$ and $B$?

We can use the fact that $\bra{n}\hat{n}\ket{n}=n$: 
\begin{eqnarray*}
\bra{n}\hat{n}\ket{n}&=&\bra{n}\hat{a}^{\dagger}\hat{a}\ket{n}
=A\bra{n}\hat{a}^{\dagger}\ket{n-1}\\ 
&=&\bra{n-1}A^{2}\ket{n-1}=A^{2}=n\\
\bra{n}\hat{n}\ket{n}&=&\bra{n}\hat{a}^{\dagger}\hat{a}\ket{n}
   =\bra{n}\hat{a}\hat{a}^{\dagger}-1\ket{n-1}\\
&=&B^{2}-1=n\\
\end{eqnarray*} 
So we see that $\hat{a}$ acts as an annihilation operator and $\hat{a}^{\dagger}$ as a creation operator:  
\begin{eqnarray}
\hat{a}^{\dagger}\ket{n}&=&\sqrt{n+1}\ket{n+1}\label{CreationEqn}\\
\hat{a}\ket{n}&=&\sqrt{n}\ket{n-1}\label{AnnihilationEqn}
\end{eqnarray}
from which we deduce that $n$ is a semi-positive number.

For $n=0$,
$\hat{a}\ket{0}=0$ and $\hat{H}\ket{0}=\frac{1}{2}\hbar
\omega\ket{0}$ representing  vacuum fluctuation energy of the lowest
eigenstate of the Hamiltonian.

The Fock states form an orthonormal set:  
\begin{eqnarray*}
\bk{n}{m}&=&\delta_{nm}\\
\sum_{n=0}^{\infty}\kb{n}{n}&=&\id
\end{eqnarray*}
And each Fock state may be built up from creation operators:  
\begin{eqnarray}
\ket{n}=\frac{\bigl( \hat{a}^{\dagger}\bigr)^{n}}{\sqrt{n!}}\ket{0}
 \end{eqnarray}

{\bf Exercise}:  show that $\ket{n}=\frac{\bigl( \hat{a}^{\dagger}\bigr)^{n}}{\sqrt{n!}}\ket{0}$.
The matrix form of the annihilation operator can be seen by using the identity operator twice:  
\begin{eqnarray*}
\hat{a}&=&\id \hat{a}\id=\sum_{n=0}^{\infty}\kb{n}{n}\hat{a}\sum_{m=0}^{\infty}\kb{m}{m}\\&=&\sum_{n=0}^{\infty}\sqrt{n+1}\kb{n}{n+1}
\end{eqnarray*}

In matrix form: 
\begin{eqnarray*}
\hat{a}=\left( \begin{array}{ccccc}
0 &\sqrt{1} &0&0&\cdots \\
0 &0&\sqrt{2}&0&\cdots \\
0&0&0&\sqrt{3}&\cdots\\
\vdots&\vdots&\vdots&\vdots&\ddots\\
\end{array}\right)
\end{eqnarray*}


In eq. (\ref{aasxpEqn}) we wrote $\hat{a}$ in terms of $\hat{x}$
and $\hat{p}$. It is often more convenient to write the annihilation
operator as a linear combination of the two Hermitian operators:
\begin{eqnarray*}
\hat{a}=\frac{\hat{Q}_{1}+i\hat{Q}_{2}}{2}
\end{eqnarray*}
where $\hat{Q}_{1}$ (equivalent to $\hat{x}$ from before) corresponds
to the in-phase component of the electric field amplitude of the
spatial-temporal mode and $\hat{Q}_{2}$ (equivalent to $\hat{p}$ from
before) corresponds to the out-of-phase component.
 
This gives the commutation relation $[\hat{Q}_{1},\hat{Q}_{2}]=2i$ and
the Heisenberg uncertainty relation
\begin{eqnarray*}
\Delta \hat{Q}_{1}\Delta \hat{Q}_{2}\ge 1
\end{eqnarray*}

With the annihilation and creation operators defined in this way it is
instructive to look at the position and momentum wave functions for
Fock states. First note that $\psi(Q_{1})=\bk{Q_{1}}{\psi}$ and
$\tilde{\psi}(Q_{2})=\bk{Q_{2}}{\psi}$, where we know that
$\hat{Q_{1}}\ket{Q_{1}}=Q_{1}\ket{Q_{1}}$ and
$\hat{Q_{2}}\ket{Q_{2}}=Q_{2}\ket{Q_{2}}$. Also note the usual
normalisation rules: $\bk{Q_{1}}{Q_{1}^{'}}=\delta(Q_{1}-Q_{1}^{'})$
and $\bk{Q_{2}}{Q_{2}^{'}}=\delta(Q_{2}-Q_{2}^{'})$, $\delta(x-x^{'})$
being the Dirac Delta function.

When we look at the position and momentum wave functions for the vacuum state we have \cite{Leonhardt} 
\begin{eqnarray*}
\bk{Q_{1}}{0}&=&\psi_{0}(Q_{1})=\pi^{-\frac{1}{4}} \exp\bigl(-\frac{Q_{1}^{2}}{4}\bigr)\\
\bk{Q_{2}}{0}&=&\tilde{\psi}_{0}(Q_{2})=\pi^{-\frac{1}{4}} \exp\bigl(-\frac{Q_{2}^{2}}{4}\bigr)
\end{eqnarray*}
And in general for a Fock state $\ket{n}$ we have
\begin{eqnarray*}
\bk{Q_{1}}{n}=\psi_{n}(Q_{1})=\frac{H_{n}(\frac{Q_{1}}{\sqrt{2}})}{\sqrt{2^{n}n!}\sqrt{\pi}}\exp\bigl(-\frac{Q_{1}^{2}}{4}\bigr)
\end{eqnarray*}
where $H_{n}(x)$ is the Hermite polynomials. 

\subsubsection{Minimum Uncertainty States}

The Heisenberg Uncertainty Principle:  for two non-commuting observables $\hat{A}$ and $\hat{B}$, the uncertainty in each is given by the relation:
\begin{eqnarray*}
\Delta \hat{A} \Delta \hat{B} \ge \frac{1}{2} |<[\hat{A},\hat{B}]>|
\end{eqnarray*}
where $(\Delta\hat{A})^{2}=\langle\hat{A}^{2}\rangle-\langle \hat{A}\rangle^{2}$. 

For example, consider the case when  $\hat{A}=\hat{x}$ and $\hat{B}=\hat{p}$. The uncertainty principle becomes the well known result:  $\Delta\hat{x}\Delta\hat{p}\ge \frac{\hbar}{2}$

A coherent state $\ket{\alpha}$ is a minimum uncertainty state. We can define it as the eigenstate of an annihilation operator:  $\hat{a}\ket{\alpha}=\alpha \ket{\alpha}$
Alternatively, we can define it with the displacement operator $D(\alpha)$:    
\begin{eqnarray*}
\ket{\alpha}&=&D(\alpha)\ket{0}=\exp(\alpha \hat{a}^{\dagger}-\alpha^{*}\hat{a})\ket{0}\\
&=&\exp(-\frac{|\alpha|^{2}}{2})\exp(\alpha \hat{a}^{\dagger})\ket{0}
\end{eqnarray*}

{\bf Exercise}:  show the last line for the above coherent state definition is correct using the Campbell-Baker-Hausdorff operator identity:  
\begin{eqnarray*}
e^{\hat{A}+\hat{B}}=e^{\hat{A}}e^{\hat{B}}e^{-\frac{1}{2}[\hat{A},\hat{B}]}
\end{eqnarray*}  
for the case when $[\hat{A},[\hat{A},\hat{B}]]=[\hat{B},[\hat{A},\hat{B}]]=0$. 

{\bf Exercise}:  show that $\hat{a}D(\alpha)\ket{0}=\alpha D(\alpha)\ket{0}$. 

If we work out $\Delta\hat{Q_{1}}\Delta\hat{Q_{2}}$ for a coherent state we find that $\Delta\hat{Q_{1}}=\Delta\hat{Q_{2}}=1$, showing that coherent states are minimum uncertainty states, since $\Delta\hat{Q_{1}}\Delta\hat{Q_{2}}=1$. 

Coherent states are not orthogonal since 
\begin{eqnarray*}
\bk{\beta}{\alpha}=\exp[-\frac{1}{2}\bigl(|\alpha|^{2}+|\beta|^{2}\bigr)+\alpha\beta^{*}]
\end{eqnarray*}
and as such form an over complete set
\begin{eqnarray*}
\int \kb{\alpha}{\alpha}d^{2}\alpha=\pi
\end{eqnarray*}

If we look at the wave functions for a coherent state ($\bk{Q_{1}}{\alpha}=\psi_{\alpha}(Q_{1})$ and $\bk{Q_{2}}{\alpha}=\tilde{\psi}_{\alpha}(Q_{2})$) \cite{Leonhardt} we find that  
\begin{eqnarray*}
\psi_{\alpha}(Q_{1})&=&\pi^{-\frac{1}{4}}\exp\Bigl( -\frac{(Q_{1}-q_{0})^{2}}{4}+\frac{i}{4}p_{0}Q_{1}-\frac{i}{8}p_{0}q_{0}\Bigr)\\
\tilde{\psi}_{\alpha}(Q_{2})&=&\pi^{-\frac{1}{4}}\exp\Bigl( -\frac{(Q_{2}-p_{0})^{2}}{4}-\frac{i}{4}q_{0}Q_{2}+\frac{i}{8}p_{0}q_{0}\Bigr)
\end{eqnarray*}
where we have decomposed the complex amplitude $\alpha$ into its real and imaginary parts:  $\alpha=\frac{1}{2}\bigl(q_{0}+ip_{0}\bigr)$.
From this we see that a coherent state is a Gaussian in $Q_{1}, Q_{2}$ phase space.


Squeezed states are another type of minimum uncertainty state. Squeezed state have one of $\Delta \hat{Q_{1}}$ or $\Delta \hat{Q_{2}}$ ``squeezed'' below 1 at the expense of the other:  either $\Delta \hat{Q_{1}}<1<\Delta \hat{Q_{2}}$ or $\Delta \hat{Q_{2}}<1<\Delta \hat{Q_{1}}$ such that $\Delta\hat{Q_{1}}\Delta\hat{Q_{2}}=1$ is always true. 

We can define squeezed states with the Squeeze operator 
\begin{eqnarray*}
S(\epsilon)=\exp{\Bigl(\frac{1}{2}\epsilon^{*}\hat{a}^{2}-\frac{1}{2} \epsilon \bigl(\hat{a}^{\dagger}\bigr)^{2}\Bigr)}
\end{eqnarray*}
where $\epsilon = r e^{i \varphi}$. 

If we define a rotated complex plane:  $\hat{P_{1}}+i \hat{P_{2}}=\bigl( \hat{Q_{1}}+i \hat{Q_{2}}\bigr)e^{-\frac{i}{2}\varphi}$ then the uncertainties can be written as $\Delta \hat{P_{1}}=e^{-r}$ and $\Delta \hat{P_{2}}=e^{r}$. Squeezed coherent states are defined to be $\ket{\alpha,\epsilon}=D(\alpha)S(\epsilon)\ket{0}$. 

If we look at the wave functions for a squeezed coherent state \cite{Leonhardt} we see that $\bra{Q_{1}}D(\alpha)S(\epsilon)\ket{0}$: 
\begin{eqnarray*}
\pi^{-\frac{1}{4}}e^{\frac{\epsilon}{2}}\exp\Bigl( -e^{2\epsilon}\frac{(Q_{1}-q_{0})^{2}}{4}+\frac{i}{4}Q_{2}Q_{1}-\frac{i}{8}p_{0}q_{0}\Bigr)
\end{eqnarray*}
where we have again decomposed the complex amplitude $\alpha$ into its real and imaginary parts:  $\alpha=\frac{1}{2}\bigl(q_{0}+ip_{0}\bigr)$.
As with the coherent state, we see that a squeezed coherent state is a Gaussian in $Q_{1}, Q_{2}$ phase space.

\subsection{Linear Optics}

What does ``Linear Optics'' mean? 

An optical component is said to be linear if its output modes $\hat{b}^{\dagger}_{j}$ 
are a linear combination of its input modes $\hat{a}^{\dagger}_{j}$: 
\begin{eqnarray*}
\hat{b}^{\dagger}_{j}=\sum_{k}M_{jk}\hat{a}^{\dagger}_{k}
\end{eqnarray*}

Linear optical components are made up of phase shifters and beam
splitters. A phase shifter is defined by the transformation
$U(P_{\phi}): \ket{n}\to e^{in\phi}\ket{n}$. That is,
$\bigl(\hat{a}_{l}^{\dagger}\bigr)^{n}\ket{0}\to
e^{in\phi}\bigl(\hat{a}_{l}^{\dagger}\bigr)^{n}\ket{0}$.

The Hamiltonian for a phase shifter is given by the number operator:
$U=e^{i\phi
  \hat{H}}=e^{i\phi\hat{n}}=e^{i\phi\hat{a}^{\dagger}\hat{a}}$. Using
this see we that $e^{i\phi\hat{a}^{\dagger}\hat{a}}\ket{n}=e^{in\phi
}\ket{n}$ {\bf Exercise:} show that the phase operator takes a
coherent state $\ket{\alpha}$ to $\ket{e^{i\phi}\alpha}$.

A beam splitter is defined by the transformation matrix 
\begin{eqnarray}
U=\left( \begin{array}{cc} \cos\theta & -e^{i\phi}\sin\theta  \\e^{-i\phi}\sin\theta & \cos\theta  	\end{array} \right) 
\label{beamsplitterEqn}
\end{eqnarray}
where the input modes are related to the output modes via
$a\dg_{l}\ket{0}\rightarrow\sum_{m}U_{ml}a\dg_{m}\ket{0}$ 
{\bf Exercise}: The Hamiltonian for a beam splitter can be given by
$\hat{H}=e^{i\phi}\hat{a}^{\dagger}\hat{b}+e^{-i\phi}\hat{a}\hat{b}^{\dagger}$,
check that this is Hermitian.

The unitary transformation for a beam splitter then looks like
$\exp\Bigl(-i\theta
\bigl(e^{i\phi}\hat{a}_{k}^{\dagger}\hat{a}_{l}+e^{-i\phi}\hat{a}_{k}\hat{a}_{l}^{\dagger}\bigr)\Bigr)$,
where $k$ and $l$ are the two modes being acted upon by the beam
splitter.

Consider we have the state $\ket{mn}_{12}$ incident on a beam
splitter. The output would look like:
\begin{eqnarray*}
\ket{mn}_{12}=\frac{\bigl(\hat{a}_{1}^{\dagger}\bigr)^{m}}
{\sqrt{m!}}\frac{\bigl(\hat{a}_{2}^{\dagger}\bigr)^{n}}{\sqrt{n!}}\ket{00}
&\to& \frac{1}{\sqrt{m!}}\Bigl(\sum_{i}U_{i1}\hat{a}_{i}^{\dagger}\Bigr)^{m}
 \frac{1}{\sqrt{n!}}\Bigl(\sum_{i}U_{j2}\hat{a}_{j}^{\dagger}\Bigr)^{n}\ket{00}\\
&=& \frac{1}{\sqrt{m!n!}}
\Bigl( \hat{a}_{1}^{\dagger}\cos\theta
    +\hat{a}_{2}^{\dagger}e^{-i\phi}\sin\theta\Bigr)^{m}
\Bigl(\hat{a}_{1}^{\dagger}e^{-i\phi}\sin\theta 
   +\hat{a}_{2}^{\dagger}\cos\theta\Bigr)^{n}\ket{00}
\end{eqnarray*}

For example:  
\begin{eqnarray*}
&\ket{10}& \rightarrow \cos\theta \ket{10} + e^{-i\phi}\sin\theta
\ket{01}\\
&\ket{01}& \rightarrow -e^{i\phi}\sin\theta \ket{10} +  \cos\theta \ket{01}\\
&\ket{11}& \rightarrow -\sqrt{2}e^{i\phi}\cos\theta \sin\theta
\ket{20} + (\cos^{2}\theta -\sin^{2}\theta)\ket{11}\\
&&\hspace{1cm}+\sqrt{2}e^{-i\phi}\cos\theta \sin\theta \ket{02}\\
&\ket{20}& \rightarrow \cos^{2}\theta \ket{20} +
\sqrt{2}e^{-i\phi}\cos\theta \sin\theta
\ket{11} + e^{-2i\phi}\sin^{2}\theta \ket{02}\\
&\ket{02}& \rightarrow e^{2i\phi}\sin^{2}\theta \ket{20} -
\sqrt{2}e^{i\phi}\cos\theta \sin\theta
\ket{11} + \cos^{2}\theta \ket{02}
\end{eqnarray*} 
{\bf Exercise}:  confirm these examples.

\subsection{Previous Suggestions With Optics}

Photons are an ideal means for communication, they travel fast
and only interact weakly with their surrounding and thus
have a long decoherence time. The problem with using solely optics for quantum
computation is how to get photons to interact. It is relatively easy
to achieve single qubit gates, but how do we establish two qubit
gates? This isn't the only problem, we also have to achieve two qubit
gates that are efficient in resources so our system will be
scalable. One way to make photons interact is to induce a cross phase
modulation between two optical modes. One example of this is the Kerr
effect.

\subsubsection{Quantum Optical Fredkin Gate}

The first proposal for quantum computation was the quantum optical
Fredkin gate \cite{Yamamoto88, Milburn89}. The gate was realised
with single photon optics using the Kerr Effect. A Fredkin
gate is a three qubit gate that acts as a controlled swap. Its truth
table is given below.

\begin{table}[h]
\begin{tabular}{cccccc}
\hline 
$c_{i}$ & $a_{i}$&$b_{i}$&$c_{o}$&$a_{o}$&$b_{o}$\\ 
\hline 
0&0&0&0&0&0\\
0&0&1&0&0&1\\
0&1&0&0&1&0\\
0&1&1&0&1&1\\
1&0&0&1&0&0\\
1&0&1&1&1&0\\
1&1&0&1&0&1\\
1&1&1&1&1&1\\
\end{tabular}
\caption{\footnotesize The logic for a Fredkin gate. Here subscript $i$ refers to the input qubits, $o$ refers to the output qubits and $c$ is the control qubit. }
\end{table} 

In this scheme, logical 0 corresponds to the vacuum mode $\ket{0}$ and
logical 1 is the 1 photon Fock state $\ket{1}$.

The optical Kerr effect is defined by a material with an intensity dependent refractive index, that is, a nonlinear crystal that has an index of refraction $n$ proportional to the total intensity $I$:  $n=n_{0}+n_{2}E^{2}=n_{0}+n_{2}I$, where $n_{0}$ is the normal refractive index and $n_{2}$ is the correction term necessary for Kerr materials. 

The Kerr Hamiltonian is of the form 
\begin{eqnarray*}
H_{I}=-\hbar \chi\hat{a}^{\dagger}_{1}\hat{a}_{1}\hat{a}^{\dagger}_{2}\hat{a}_{2}
\end{eqnarray*}
where $\chi$ is a coupling constant which depends upon the third-order
non-linear susceptibility for the optical Kerr effect and
$\hat{a}_{1}$ and $\hat{a}_{2}$ are the annihilation modes for the
input light.

The optical setup for the Fredkin gate first proposed in
\cite{Milburn89} is shown below in fig. \ref{MilburnPic}. In this
setup we have a Mach-Zehnder interferometer with a Kerr media in
either arm. In the top arm the Kerr media has the control beam
incident on it.

\begin{figure}[h!]
\begin{center}
\includegraphics[width=0.6\textwidth]{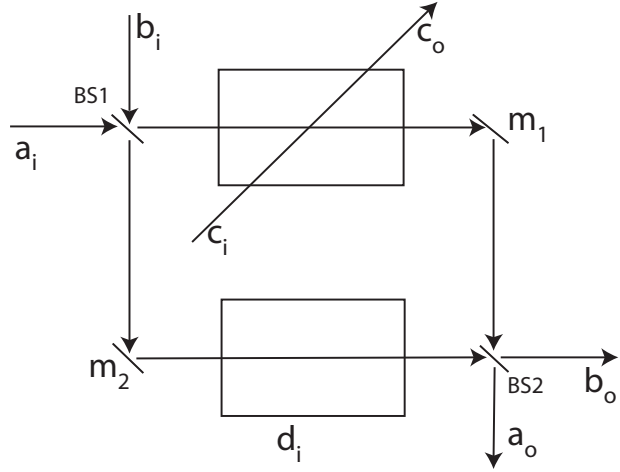}
\end{center}
\caption{\footnotesize The original quantum Fredkin gate \cite{Milburn89}. }
\label{MilburnPic}
\end{figure}
The beam splitter on the left side, BS1, has $\theta=\frac{\pi}{4}, \phi=0$
and the beam splitter on the right hand side, BS2, has $\theta=\frac{\pi}{4}, \phi=\pi$. The effect
of the Kerr media is to act the unitary operator 
$U=\exp{\bigl(-i\frac{t}{\hbar} H_{I}\bigr)}
=\exp{\bigl(i\epsilon\hat{a}^{\dagger}\hat{a}\hat{c}^{\dagger}\hat{c}\bigr)}$
in the top arm and
$U=\exp{\bigl(i\epsilon\hat{b}^{\dagger}\hat{b}\hat{d}^{\dagger}\hat{d}\bigr)}$
in the bottom arm, where $\epsilon=t\chi$.

For example, consider the input state $\ket{cab,d}=\ket{010,0}$
\begin{eqnarray*}
\ket{010,0}&&\rightarrow \frac{1}{\sqrt{2}}\bigl( \ket{010,0}+\ket{001,0}\bigr)\\
&&\rightarrow \frac{1}{\sqrt{2}}e^{i \epsilon\hat{c}^{\dagger}\hat{c}\hat{a}^{\dagger}\hat{a}}e^{i \epsilon\hat{b}^{\dagger}\hat{b}\hat{d}^{\dagger}\hat{d}}\bigl( \ket{010,0}+\ket{001,0}\bigr)\\
&&= \frac{1}{\sqrt{2}}
\bigl(e^{i \epsilon\hat{c}^{\dagger}\hat{c}\hat{a}^{\dagger}\hat{a}} \ket{010,0}
+e^{i \epsilon\hat{b}^{\dagger}\hat{b}\hat{d}^{\dagger}\hat{d}}\ket{001,0}\bigr)\\
&& =\frac{1}{\sqrt{2}}\bigl( \ket{010,0}+\ket{001,0}\bigr)\rightarrow \ket{010,0}
\end{eqnarray*} 
where mode $d$ is always in the vacuum mode $\ket{0}$ and remember that 
$e^{i \phi \hat{n}}\ket{m}=e^{i m \phi }\ket{m}$.

Next consider the input $\ket{101,0}$
\begin{eqnarray*}
\ket{101,0}&&\rightarrow \frac{1}{\sqrt{2}}\bigl( \ket{101,0}-\ket{110,0}\bigr)\\
&&\rightarrow\frac{1}{\sqrt{2}}
\bigl(-e^{i \epsilon\hat{c}^{\dagger}\hat{c}\hat{a}^{\dagger}\hat{a}} \ket{110,0}
+e^{i \epsilon\hat{b}^{\dagger}\hat{b}\hat{d}^{\dagger}\hat{d}}\ket{101,0}\bigr)\\
&&\rightarrow \frac{1}{\sqrt{2}}\bigl(-e^{i \epsilon} \ket{110,0}+\ket{101,0}\bigr)\\
&&\rightarrow \frac{1}{2}
\bigl( -e^{i \epsilon} \ket{110,0}+e^{i \epsilon} \ket{101,0}
        +\ket{101,0}+\ket{110,0}\bigr).
\end{eqnarray*}
If we choose $\epsilon=\pi$ this gives $\ket{110,0}$. With this choice
of $\epsilon$ we can show that all the logic gates are as in the
Fredkin gate logic table.

There are two major problems with this scheme: (1) it is difficult to
achieve the high non-linearities, especially those required for 
$\epsilon=\pi$ phase change; (2) at high non-linearities the crystal exhibits
other detrimental affects, such as absorption.

\subsubsection{Cavity Quantum Electrodynamics}

Using non-linear crystals is not the only method to induce a coupling
between optical modes. In \cite{Turchette95} Turchette
\textit{et. al.} used cavity quantum electrodynamics to induce a two
qubit interaction between photons. To do this single atoms in a
Fabry-Perot cavity were used.

We start with the Jaynes-Cummings Hamiltonian for the interaction of
photons with a two level atom:

\begin{eqnarray}
  H=\frac{1}{2}\hbar \omega_{0} (-Z)
   +\hbar\omega \hat{a}^{\dagger}\hat{a}
   +g\bigl(\hat{a}^{\dagger}\sigma_{-}+\hat{a}\sigma_{+}\bigr)
\end{eqnarray}
where $\omega_{0}$ is the frequency of the atom, $\omega$ the frequency
of the field and $g$ is the coupling constant between the field and
atom.

The first term in the above Hamiltonian is the atom energy term
$H_{\text{atom}}=-\frac{1}{2}\hbar \omega_{0} Z$. We have a two level
atom and we say the energy of the top level ($\ket{1}$) is
$\frac{1}{2}\hbar\omega_{0}$ and that of the bottom level
$\ket{0}$ is $-\frac{1}{2}\hbar\omega_{0}$. From the difference in energy we have $H_{\text{atom}}=\frac{1}{2}\hbar\omega_{0}\bigl(\kb{1}{1}-\kb{0}{0}\bigr)=-\frac{1}{2}\hbar \omega_{0} Z$, given that $Z=-(\kb{1}{1}-\kb{0}{0})$. The $H_{\text{field}}=\hbar\omega \hat{a}^{\dagger}\hat{a}$ term comes from the harmonic oscillator describing light (and we have dropped the
$\frac{1}{2}\hbar\omega$ term as it plays no role here) Finally, there
is the interaction term $H_{I}$ which shows that an atom in the state
$\ket{1}$ can decay to give a photon, or a photon can excite the atom
from a state $\ket{0}$. $\sigma_{\pm}$ are the usual raising and
lowering operators,
$\sigma_{+}=\kb{1}{0}=\frac{1}{2}\bigl(X+iY\bigr)$,
$\sigma_{-}=\kb{0}{1}=\frac{1}{2}\bigl(X-iY\bigr)$. The terms
$\hat{a}^{\dagger}\sigma_{+}$ or $\hat{a}\sigma_{-}$ are dropped from
this Hamiltonian by the so called rotating wave approximation.

We can rewrite the Jaynes-Cummings Hamiltonian in the following form,
as done in~\cite{Nielsen}

\begin{eqnarray}
H=\hbar\omega N+\delta Z
   +g\bigl(\hat{a}^{\dagger}\sigma_{-}+\hat{a}\sigma_{+}\bigr)
\end{eqnarray}
where $N=\hat{a}^{\dagger}\hat{a}+\frac{1}{2}Z$ and
$\delta=\frac{1}{2}\hbar\bigl( \omega_{0}-\omega\bigr)$. We may
neglect $N$ from any further analysis as it leads to a constant phase
factor that we can neglect. We thus have a Hamiltonian that is
reduced to:

\begin{eqnarray}
H=\delta Z+g\bigl(\hat{a}^{\dagger}\sigma_{-}+\hat{a}\sigma_{+}\bigr)
\label{CavityHEqn}
\end{eqnarray}

The goal is to utilise a single atom to obtain a non-linear
interaction between two photons. The quantum information will be
represented by orthogonal polarisation photon states, using a cavity
with a single atom to provide the non-linear interaction between the
photons. We can modify the Hamiltonian (\ref{CavityHEqn}) to incorporate
two photons of slightly different frequencies and take photon
polarisation into account. A three level atom that is excited to a
different energy level (of approximately the same energy) by the two
orthogonal polarisation, a so called $V$ type atom, suffices for this.

Turchette \textit{at. al.} used Cesium atoms and two photons, one
circularly polarised (mode $a$) and the other of slightly different
frequency also circularly polarised (mode $b$), both generated from
weak laser pulses. The logic was represented by the two orthogonal
polarisation modes $\ket{1^{\pm}}_{a}\ket{1^{\pm}}_{b}$. The phases
induced by interaction with the atom were then measured using
heterodyne measurement.

The following logical transformations were measured:
\begin{eqnarray*}
\ket{1^{-}}_{a}\ket{1^{-}}_{b}&\to &\ket{1^{-}}_{a}\ket{1^{-}}_{b}\\
\ket{1^{+}}_{a}\ket{1^{-}}_{b}&\to &e^{i\phi_{a}}\ket{1^{+}}_{a}\ket{1^{-}}_{b}\\
\ket{1^{-}}_{a}\ket{1^{+}}_{b}&\to &e^{i\phi_{b}}\ket{1^{-}}_{a}\ket{1^{+}}_{b}\\
\ket{1^{+}}_{a}\ket{1^{+}}_{b}&\to &
                     e^{i(\phi_{a}+\phi_{b}+\Delta)}\ket{1^{+}}_{a}\ket{1^{+}}_{b}
\end{eqnarray*}
where $\phi_{a}\approx(17.5\pm1)^{o}$, $\phi_{b}\approx(12.5\pm1)^{o}$
and $\Delta\approx(16\pm3)^{o}$.

The beauty of this interaction is that $\ket{1^{+}1^{+}}$ goes to
$e^{\bigl(\varphi_{a}+\varphi_{b}+\Delta\bigr)}\ket{1^{+}1^{+}}$,
where intuitively one might expect that $\ket{1^{+}1^{+}}$ would go to
$e^{\bigl(\varphi_{a}+\varphi_{b}\bigr)}\ket{1^{+}1^{+}}$. This is the
photon non-linearity. If $\Delta$ were 0, the system would be linear
and the photons would not interact. This is the key for implementing
quantum logic with these methods. Unfortunately it is much harder to get higher non linearities such as $\Delta= 90^{o}$ and does not appears to  be a promising approach when compounding many gates.

\subsection{Progress with Linear Optics}

Another direction to devise quantum information processor is to focus
on linear optics alone.  In this section we show what progress was made 
in this direction before the proposal by Knill \textit{et. al.} \cite{KLM}.

\subsubsection{Decomposition of unitaries}

Reck \textit{et. al.} showed that we can break {\it any}
unitary into a set of linear optical components \cite{Reck98}. The unitary
transformation considered is acting on the creation
operators, so one is not able to immediately apply this method to say
to the CNOT unitary.

A unitary operator $U$ of dimension $N$ can be decomposed as follows:  
\begin{eqnarray}
U=\Bigl(T_{N,N-1}\cdot T_{N,N-2}\cdot T_{N,N-3}
\cdots T_{N,1}\cdot T_{N-1,N-2}\cdot T_{N-1,N-3}\cdots T_{2,1}D\Bigr)^{-1}
\end{eqnarray}
where $T_{p,q}$ is the $N$ dimensional identity with the $\{ p,q\}$
elements replaced with the beam splitter matrix (\ref{beamsplitterEqn})
and $D$ is an $N\times N$ matrix with phases on the diagonal.

\begin{figure}[h!]
\begin{center}
\includegraphics[width=0.8\textwidth]{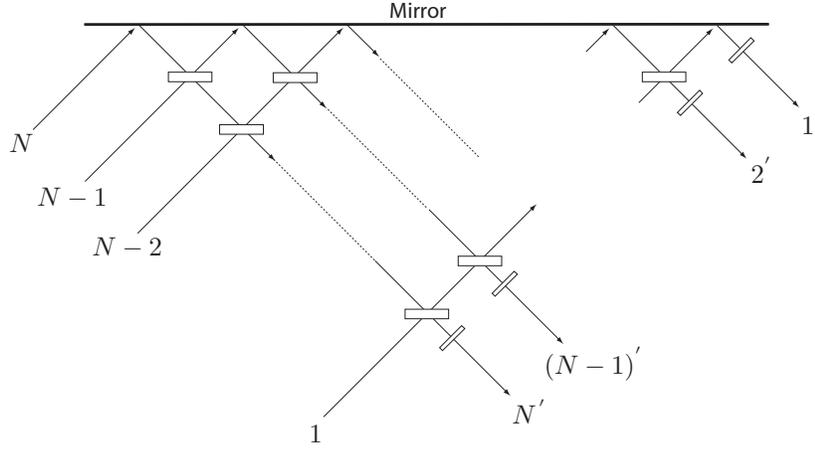}
\end{center}
\caption{\footnotesize General linear optical network for a unitary matrix $U$.}
\label{ReckPic}
\end{figure}

The general linear optical network for a unitary matrix $U$ is a
triangular array of beam splitters and phase shifters, shown in fig.
\ref{ReckPic}.

As an example, what does the linear optical circuit for the following
unitary look like?
\begin{eqnarray*}
U=\left( \begin{array}{ccc}
1-\sqrt{2} &\frac{1}{\sqrt{\sqrt{2}}} &\sqrt{\frac{3}{\sqrt{2}}-2}\\
\frac{1}{\sqrt{\sqrt{2}}} &\frac{1}{2}&\frac{1}{2}-\frac{1}{\sqrt{2}}\\
\sqrt{\frac{3}{\sqrt{2}}-2}&\frac{1}{2}-\frac{1}{\sqrt{2}}&\sqrt{2}-\frac{1}{2}\\
\end{array}\right)
\end{eqnarray*}
{\bf Exercise}:  check that this is unitary. 

Here we have a unitary matrix with N=3, so we can say that 
\begin{eqnarray*}
U=\Bigl( T_{3,2}\cdot T_{3,1}\cdot T_{2,1}\cdot D\Bigr)^{-1}=D^{\dagger}\cdot T_{2,1}^{\dagger}\cdot T_{3,1}^{\dagger}\cdot T_{3,2}^{\dagger}
\end{eqnarray*}

We find that 
\begin{eqnarray*}
T_{3,2}=\left( \begin{array}{ccc}
1&0&0\\
0&\cos\theta_{1}&e^{-i \phi_{1}}\sin\theta_{1}\\
0&-e^{i \phi_{1}}\sin\theta_{1}&\cos\theta_{1}\\
\end{array}\right)
\end{eqnarray*}
\begin{eqnarray*}
T_{3,1}=\left( \begin{array}{ccc}
\cos\theta_{2}&0&e^{-i \phi_{2}}\sin\theta_{2}\\
0&1&0\\
-e^{i \phi_{2}}\sin\theta_{2}&0&\cos\theta_{2}\\
\end{array}\right)
\end{eqnarray*}
\begin{eqnarray*}
T_{2,1}=\left( \begin{array}{ccc}
\cos\theta_{3}&e^{-i \phi_{3}}\sin\theta_{3}&0\\
-e^{i \phi_{3}}\sin\theta_{3}&\cos\theta_{3}&0\\
0&0&1\\
\end{array}\right)
\end{eqnarray*}

and 
\begin{eqnarray*}
D=\left( \begin{array}{ccc}
e^{i\phi_{4}}&0&0\\
0&e^{i\phi_{5}}&0\\
0&0&e^{i\phi_{6}}\\
\end{array}\right)
\end{eqnarray*}
{\bf Exercise}:  show that a solution to this is given by the angles: 
$\theta_{1}=12.8^{o}, \phi_{1}=\pi$, $\theta_{2}=20.4^{o}, \phi_{2}=0$, $\theta_{3}=63.8, \phi_{3}=0$, $\phi_{4}=\pi, \phi_{5}=0, \phi_{6}=0$. 

{\bf Exercise}: What transformation does this implement on the input
state $\bigl(\alpha\ket{0}+\beta\ket{1}+\gamma\ket{2}\bigr)\ket{10}$?
What is the transformation if we measure $\ket{10}$ in modes 2 and 3?
This will be important later!

\subsubsection{Optical Simulation of Quantum Logic}

In \cite{Cerf98}, Cerf \textit{et. al.} proposed a scheme for quantum logic with only linear optical devices and a single photon. To simulate $n$ qubits a single photon is put into $2^{n}$ different paths. 

A $\sqrt{\text{NOT}}$ gate is given by a beam splitter with $\theta=\frac{\pi}{4}$ and $\phi=-\frac{\pi}{2}$, where 
\begin{eqnarray*}
\sqrt{\text{NOT}}=\frac{1}{\sqrt{2}}\left( \begin{array}{cc}
1&i\\
i&1\\
\end{array}\right)
\end{eqnarray*} 
\begin{eqnarray*}
\ket{0}_{q}=\ket{01}&\to&\frac{1}{\sqrt{2}}\bigl(i\ket{10}+\ket{01}\bigr)\\
\ket{1}_{q}=\ket{10}&\to&\frac{1}{\sqrt{2}}\bigl(\ket{10}+i\ket{01}\bigr)
\end{eqnarray*}

\begin{figure}[h!]
\begin{center}
\includegraphics[width=0.8\textwidth]{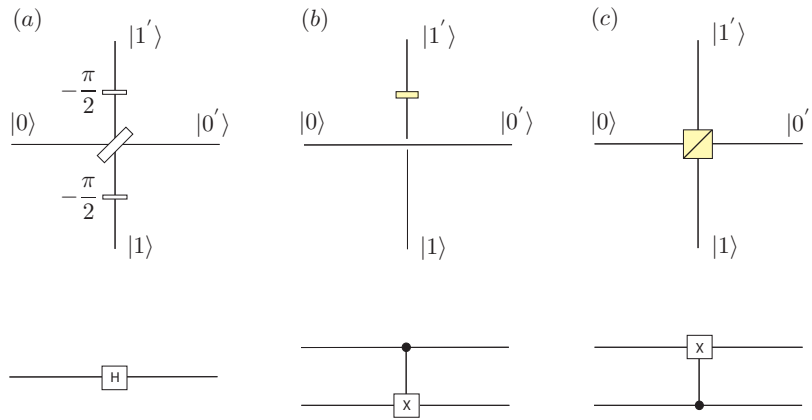}
\end{center}
\caption{\footnotesize A Hadamard, a CNOT and a reverse CNOT. }
\label{CerfPic}
\end{figure}

Three simple gate implementations proposed by Cerf \textit{et. al.} are given in fig. \ref{CerfPic}.

With these a universal set of gates is possible. 

To implement a Hadamard gate we use a  $\theta=\frac{\pi}{4},\phi=-\frac{\pi}{2}$ beam splitter and two $-\frac{\pi}{2}$ phase shifters, as in part (a) of fig. \ref{CerfPic}. 
\begin{eqnarray*}
\ket{0}_{q}&=&\ket{01}\rightarrow -i \ket{01}\rightarrow -\frac{i}{\sqrt{2}}\bigl( \ket{01}+i \ket{10}\bigr)\\
&&\rightarrow  \frac{-i}{\sqrt{2}}\bigl( \ket{01}+i(-i) \ket{10}\bigr)= \frac{-i}{\sqrt{2}}\bigl( \ket{01}+ \ket{10}\bigr)\\
\ket{1}_{q}&=&\ket{10}\rightarrow \ket{10}\rightarrow \frac{1}{\sqrt{2}}\bigl(i \ket{01}+\ket{10}\bigr)\\
&&\rightarrow  \frac{1}{\sqrt{2}}\bigl( i\ket{01}+-i \ket{10}\bigr)= \frac{i}{\sqrt{2}}\bigl( \ket{01}- \ket{10}\bigr)
\end{eqnarray*}

To implement a CNOT we encode a qubit in position and polarisation. The location is the control and the polarisation the target. For the control:  $\ket{0}_{q}=\ket{01}$, $\ket{1}_{q}=\ket{10}$. For the target:  $\ket{0}_{q}=\ket{H}$, $\ket{1}_{q}=\ket{V}$. 
The circuit for this is a polarisation rotator on the upper arm, as in part (b) of fig. \ref{CerfPic}. If a photon is present in the top arm, its polarisation will be flipped:  
\begin{eqnarray*}
\bigl(&&\alpha\ket{01}+\beta\ket{10}\bigr)\otimes \bigl( \gamma\ket{H}+\delta\ket{V}\bigr)\\
&&\rightarrow \alpha\gamma \ket{01}\ket{H}+\alpha \delta\ket{01}\ket{V}+\beta\gamma\ket{10}\ket{V}
+\beta\delta\ket{10}\ket{H}
\end{eqnarray*}

To implement a reverse CNOT we simply need a polarising beam splitter (PBS), where horizontal is reflected. As before, the location is the control and the polarisation the target. This is in part (c) of fig. \ref{CerfPic}: 
\begin{eqnarray*}
\bigl(&&\alpha\ket{01}+\beta\ket{10}\bigr)\otimes \bigl( \gamma\ket{H}+\delta\ket{V}\bigr)\\
&&\rightarrow \alpha\gamma \ket{01}\ket{H}+\alpha \delta\ket{10}\ket{V}+\beta\gamma\ket{10}\ket{H}
+\beta\delta\ket{01}\ket{V}
\end{eqnarray*}

The problem with this scheme is that $n$ qubits requires $2^{n}$ paths
which in turn requires $2^{n}-1$ beam splitters to setup. This is not
scalable. With one qubit encoded in polarisation we still need
$2^{n-1}$ optical paths.


\section{Linear Optics Quantum Compution}

In this section we will develop ideas and concepts for linear optics quantum computing. We will show that using beam splitters, phase shifters, single photon sources and detectors we will be able to simulate efficiently, i.e. using a polynomial amount of resources, an ideal quantum computer.  It is known that a device using only linear optics (beam splitters, phase shifters)  is not more powerful than a classical computer. We will see that  once we use the result of measurement and feed it back to future linear gates, we can now simulate a quantum computer.

\subsection{Assumptions in LOQC}

Before we start explaining the details of linear optics quantum
computing (LOQC), we first recall what ``linear'' means when we
say \textit{linear optics}.

For an optical element to be linear, each output mode
$\hat{b}^{\dagger}_{j}$ is a linear sum of the input modes
$\hat{a}^{\dagger}_{k}$:
\begin{eqnarray*}
\hat{b}^{\dagger}_{j}=\sum_{k}M_{jk}\hat{a}^{\dagger}_{k}
\end{eqnarray*} 
where $M$ is a unitary matrix.

We use beam splitters and phase shifter to implement linear optics:

A phase shifter is described via: $U=e^{i n \phi}$, where $n$ is the
number of photons in the mode.

\begin{figure}[h!]
\begin{center}
\includegraphics[width=0.22\textwidth]{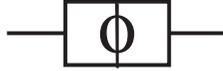}
\end{center}
\caption{\footnotesize A phase shifter of phase $\phi$.}
\label{phaseshifterPic}
\end{figure}

A beam splitter is described by:  
\begin{eqnarray}
U=\left( \begin{array}{cc} \cos\theta & -e^{i\phi}\sin\theta  \\e^{-i\phi}\sin\theta & \cos\theta  \\ 	\end{array} \right) 
\end{eqnarray}
where the input modes are related to the output modes via 
\begin{eqnarray*}
a\dg_{l}\ket{0}\rightarrow\sum_{m}U_{ml}a\dg_{m}\ket{0}
\end{eqnarray*}

\begin{figure}[h!]
\begin{center}
\includegraphics[width=0.22\textwidth]{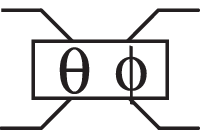}
\end{center}
\caption{\footnotesize A beam splitter of angle $\theta$ and phase $\phi$.}
\label{beamsplitterPic}
\end{figure}

The basic resources that are necessary for linear optical quantum
computing are single photon sources and detectors that can distinguish
between 0,1 and 2 photons.

\subsection{Qubits in LOQC}

In LOQC we encode qubits with dual rail logic. That is, logic that is
encoded using the physical location of a photon. Logical 0 is given by
$\ket{0}_{q}=\ket{01}_{ab}$ and logical 1 by
$\ket{1}_{q}=\ket{10}_{ab}$, where mode $a$ is one spatial mode and
mode $b$ another.

We could just as easily use photon polarisation:
$\ket{0}_{q}=\ket{H}$, $\ket{1}_{q}=\ket{V}$. We can change
the representation of the qubit from polarisation to spatial mode
using a polarising rotator and a polarising beam splitter as can be seen below 
in fig. \ref{DualToPolPic}.

\begin{figure}[h!]
\begin{center}
\includegraphics[width=0.8\textwidth]{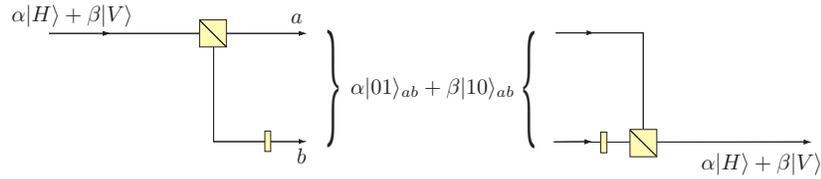}
\end{center}
\caption{\footnotesize Transforming from dual rail to polarisation encoding.}
\label{DualToPolPic}
\end{figure}

The polarisation encoding of qubits is useful to understand some of the 
experimental implementation of LOQC but for this and subsequent lectures we will be using dual rail encoding for qubits. 

\subsection{Qubit Operations}

For a physical system to be a viable candidate for quantum
computation, we need a universal set of gates, as stated in point 3 of
DiVincenzo's five criteria \cite{DiVincenzo00}, seen in 
section~\ref{sect:qunatumcomputation}. 
Such a universal set is comprised of arbitrary single qubit
gates and the CNOT operation \cite{DiVincenzo95}.

We will show that single qubit operations are easily performed using
linear elements and almost always error free. However, as we saw last
lecture, it is quite difficult to make photons interact. We will show
that with the use of single photon ancillas and photo-detection we can
make a two qubit gate (a control phase gate). 
The non-linearity will still be there, in the form of measurement.
Unfortunately this two qubit gate will not be unitary but using quantum
teleportation and quantum error correction we will be able 
to approximate a unitary operation using only a polynomial amount
of resources.

\subsection{Single qubit gates}

All single qubit gates can be implemented with just beam splitters and
phase shifters. To see this, note that any single qubit unitary, $U$,
can be decomposed into rotations about the $Z$ and $Y$ axis in the
Bloch sphere as follows \cite{Nielsen}:
\begin{eqnarray*}
U=e^{i\alpha}R_{z}(\beta)R_{y}(\gamma)R_{z}(\delta)
\end{eqnarray*}
where 
$R_{z}(\theta)=e^{-i\frac{\theta}{2}\sigma_{z}}$  and 
$R_{y}(\phi)=e^{-i\frac{\phi}{2}\sigma_{y}}$. 
{\bf Exercise}:  show that 
$R_{x}(\vartheta)=R_{z}(\frac{\pi}{2})R_{y}(-\vartheta)R_{z}(-\frac{\pi}{2})$. 

If we can show how to make arbitrary rotations about the $Z$ and $Y$
axis we can perform any arbitrary single qubit operation.

Rotations about the $Z$ axis can simply be performed using a phase shifter on the top mode of a qubit:

\begin{figure}[h!]
\begin{center}
\includegraphics[width=0.35\textwidth]{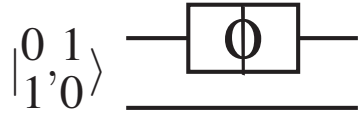}
\end{center}
\caption{\footnotesize Performs a rotation of $\phi$ about the $Z$ axis:  
$R_{Z}(\phi)$.}
\label{onequbitZPic}
\end{figure}

Consider the effect of the above circuit on the qubit
$\alpha\ket{0}_q+\beta\ket{1}_q$:
\begin{eqnarray*}
\alpha\ket{0}_q+\beta\ket{1}_q &=& \alpha\ket{01}+\beta\ket{10} \\
&\rightarrow& \alpha\ket{01}+\beta e^{i\phi}\ket{10} \\
&=& e^{i\phi/2}( e^{-i\phi/2}\alpha\ket{0}_q+e^{i\phi/2}\beta\ket{ 1}_q) \\
&=& e^{i\phi/2}e^{-i\phi Z_q/2} (\alpha\ket{0}_q+ \beta\ket{1}_q)\\
&=& e^{i\phi/2}R_{Z}(\phi) (\alpha\ket{0}_q+ \beta\ket{1}_q)\\
\end{eqnarray*}
And we see that, up to an irrelevant global phase, 
a $R_{Z}(\phi)$ has been performed. 

Rotations of $-2\theta$ about the $Y$ axis require a beam splitter of
angle $\theta$ and angle $\phi=0$:

\begin{figure}[h!]
\begin{center}
\includegraphics[width=0.25\textwidth]{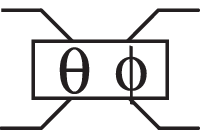}
\end{center}
\caption{\footnotesize Performs a rotation of $-2\theta$ about the $Y$
  axis: $R_{Y}(-2\theta)$.}
\label{onequbitYPic}
\end{figure}

We can see this by first remembering the effect of a beam splitter on
a photon in modes 1 and 2:
\begin{eqnarray*}
\hat{a}_{1}^{\dagger}&\rightarrow& 
\bigl(\cos(\theta)\hat{a}_{1}^{\dagger}+\sin(\theta)\hat{a}_{2}^{\dagger}\bigr) \\
\hat{a}_{2}^{\dagger}&\rightarrow& 
\bigl(-\sin(\theta)\hat{a}_{1}^{\dagger}+\cos(\theta)\hat{a}_{2}^{\dagger}\bigr) 
\end{eqnarray*}
We can consider the effect of the above circuit on the qubit 
$\alpha\ket{0}_q+\beta\ket{1}_q$:
\begin{eqnarray*}
\alpha\ket{0}_q+\beta\ket{1}_q &=& \alpha\ket{01}+\beta\ket{10} \\
&\rightarrow& \alpha\bigl(\cos(\theta)\ket{01}-\sin(\theta)\ket{10}\bigr)+\beta\bigl(\cos(\theta)\ket{10}+\sin(\theta)\ket{01}\bigr) \\
&=& \cos(\theta) \bigl(\alpha\ket{01}+\beta\ket{10}\bigr) -\sin(\theta)\bigl(\alpha\ket{10}-\beta\ket{01}\bigr)\\
&=& e^{i\theta Y_q} \bigl(\alpha\ket{0}_q+ \beta\ket{1}_q\bigr)\\
&=& R_{Y}(-2\theta)\bigl(\alpha\ket{0}_q+ \beta\ket{1}_q\bigr)\\
\end{eqnarray*}

\subsection{Two qubit gates}

Now we have to address the question of how to make a CNOT gate with just single photon ancillas and photo-detection. We will actually be looking at how we can perform a CSign, the relation between the two is apparent from the fact that $HZH=X$. The CSign gate has the action $\ket{x}_{L}\ket{y}_{L}\to e^{i\pi x\cdot y}\ket{x}_{L}\ket{y}_{L}$, $x,y\in\{0,1\}$. 
\begin{figure}[h!]
\begin{center}
\includegraphics[width=1.0\textwidth]{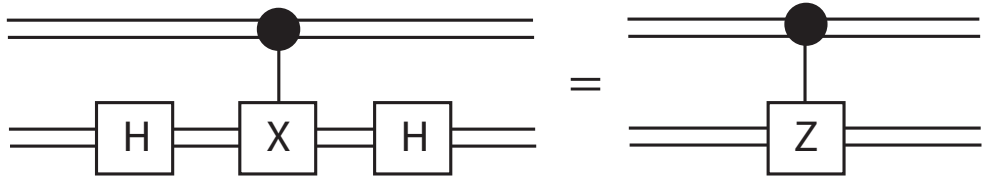}
\end{center}
\caption{\footnotesize The equivalence between a CNOT and a CSign.}
\label{CNOTHadPic}
\end{figure}

We will first show how to perform a probabilistic CSign gate on
photonic qubits. From here we will show how the CSign probability of
success can be boosted arbitrarily close to 1 using teleportation. In
the next section we will see how this could be simplified even further
by using quantum error correction.

To perform a CSign gate we need a basic transformation:
$\alpha\ket{0}+\beta\ket{1}+\gamma\ket{2}\to
\alpha\ket{0}+\beta\ket{1}-\gamma\ket{2}$. This
is called a non-linear sign shift $NS_{-1}$. 
{\bf Exercise}: why is the transformation non-linear?
 
\subsubsection{Nonlinear sign shift gate}

As the above transformation is non-linear we cannot use
linear optics alone. We find that with the use of ancilla modes and
photo-detection we can indeed perform this transformation using two
extra modes, one with a single photon and the other in vacuum as shown
in the circuit below. The input state looks like
$\bigl(\alpha\ket{0}_{1}+\beta\ket{1}_{1}+\gamma\ket{2}_{1}\bigr)\otimes
\ket{1}_{2}\otimes\ket{0}_{3}$.

\begin{figure}[h!]
\begin{center}
\includegraphics[width=0.8\textwidth]{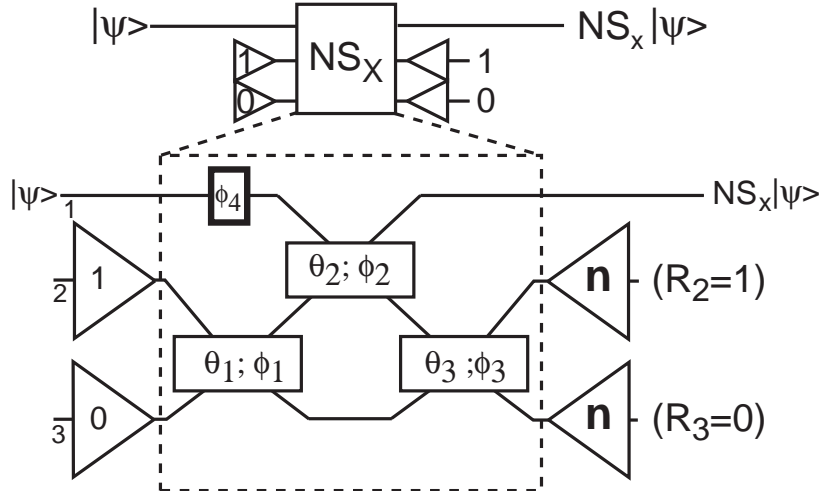}
\end{center}
\caption{\footnotesize The $NS_{-1}$ gate.}
\label{fig1KLMPic}
\end{figure}
Where $\theta_{1}=22.5^{\circ}, \phi_{1}=0^{\circ}$, $\theta_{2}=65.5302^{\circ}, 
\phi_{2}=0^{\circ}$, $\theta_{3}=-22.5^{\circ}, \phi_{3}=0^{\circ}$ and 
$\phi_{4}=\pi^{\circ}$. 

The action of this circuit is to apply the following unitary matrix on the input modes: 
\begin{eqnarray*}
U\left(\begin{array}{ccc} \hat{a}_{1}^{\dagger}\\ \hat{a}_{2}^{\dagger}\\ \hat{a}_{3}^{\dagger}\end{array}\right)
=\left( \begin{array}{ccc}
1-\sqrt{2} &\frac{1}{\sqrt{\sqrt{2}}} &\sqrt{\frac{3}{\sqrt{2}}-2}\\
\frac{1}{\sqrt{\sqrt{2}}} &\frac{1}{2}&\frac{1}{2}-\frac{1}{\sqrt{2}}\\
\sqrt{\frac{3}{\sqrt{2}}-2}&\frac{1}{2}-\frac{1}{\sqrt{2}}&\sqrt{2}-\frac{1}{2}\\
\end{array}\right)
\left(\begin{array}{ccc} \hat{a}_{1}^{\dagger}\\ \hat{a}_{2}^{\dagger}\\ \hat{a}_{3}^{\dagger}\end{array}\right)
\end{eqnarray*}
Does this look familiar? Remember lecture 1. 

When we measure a single photon in mode 2 and vacuum in mode 3 we have the state $\alpha\ket{0}_{1}+\beta\ket{1}_{1}-\gamma\ket{2}_{1}$ output in mode 1. The probability of measuring a single photon in mode 2 and vacuum in mode 3 is $\frac{1}{4}$. 

{\bf Exercise}:  show that this circuit indeed performs the transformation $\alpha\ket{0}+\beta\ket{1}+\gamma\ket{2}\to\alpha\ket{0}+\beta\ket{1}-\gamma\ket{2}$ when mode 2 and 3 are measured in $\ket{10}$ and that this succeeds with probability $1/4$.

\subsubsection{Controlled Sign gate}

With the use of two $NS_{-1}$ gates we can make a CSign gate, as seen in the fig. below. 
\begin{figure}[h!]
\begin{center}
\includegraphics[width=0.5\textwidth]{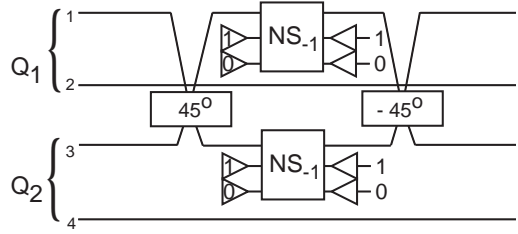}
\end{center}
\caption{\footnotesize A probabilistic CSign gate. The probability of success $\frac{1}{16}$.}
\label{fig2KLMPic}
\end{figure}

Lets work through a general example, say we start with the input state $Q_{1}=\alpha\ket{0}_q+\beta\ket{1}_q=\alpha\ket{01}+\beta\ket{10}$ in modes 1 and 2 and $Q_{2}=\gamma\ket{0}_q+\delta\ket{1}_q=\gamma\ket{01}+\delta\ket{10}$ in modes 3 and 4.

After the first beam splitter we have 
\begin{eqnarray*}
&&\alpha\gamma\ket{0101}+\alpha\delta\ket{0110}+\beta\gamma\ket{1001}+\beta\delta\ket{1010}\\ 
&\to& \alpha\gamma\ket{0101}-\frac{\alpha\delta}{\sqrt{2}}\ket{1100}+\frac{\alpha\delta}{\sqrt{2}}\ket{0110}+\frac{\beta\gamma}{\sqrt{2}}\ket{1001}\\ 
&&\,\,+\frac{\beta\gamma}{\sqrt{2}}\ket{0011}-
   \frac{\beta\delta}{\sqrt{2}}\ket{2000}+
   \frac{\beta\delta}{\sqrt{2}}\ket{0020}
\end{eqnarray*}

\begin{figure}[h!]
\begin{center}
\includegraphics[width=0.5\textwidth]{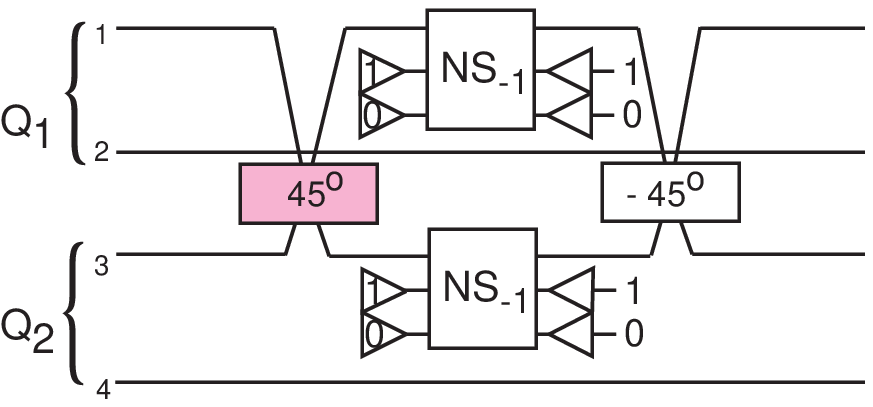}
\end{center}
\caption{\footnotesize Step 1 of a probabilistic CSign gate.}
\label{fig2KLMstep1Pic}
\end{figure}

After the two $NS_{-1}$ gates we have 
\begin{eqnarray*}
&\to &  \alpha\gamma\ket{0101}-\frac{\alpha\delta}{\sqrt{2}}\ket{1100}+\frac{\alpha\delta}{\sqrt{2}}\ket{0110}+\frac{\beta\gamma}{\sqrt{2}}\ket{1001}\\ &&\,\,+\frac{\beta\gamma}{\sqrt{2}}\ket{0011}+\frac{\beta\delta}{\sqrt{2}}\ket{2000}-\frac{\beta\delta}{\sqrt{2}}\ket{0020}\\
\end{eqnarray*}

\begin{figure}[h!]
\begin{center}
\includegraphics[width=0.5\textwidth]{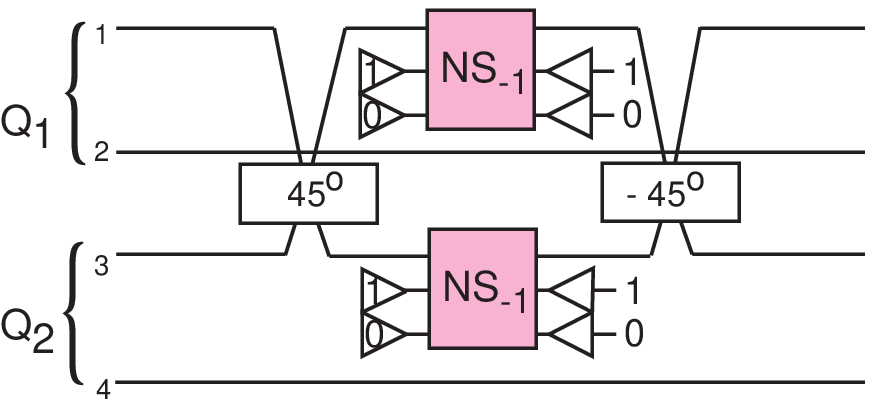}
\end{center}
\caption{\footnotesize Step 2 of a probabilistic CSign gate.}
\label{fig2KLMstep2Pic}
\end{figure}

After the final beam splitter we have 
\begin{eqnarray*}
&\to &  \alpha\gamma\ket{0101}+\alpha\delta\ket{0110}+\beta\gamma\ket{1001}-\beta\delta\ket{1010}\\
&= &  \alpha\gamma\ket{00}_q+\alpha\delta\ket{01}_q+\beta\gamma\ket{10}_q-\beta\delta\ket{11}_q
\end{eqnarray*}

\begin{figure}[h!]
\begin{center}
\includegraphics[width=0.5\textwidth]{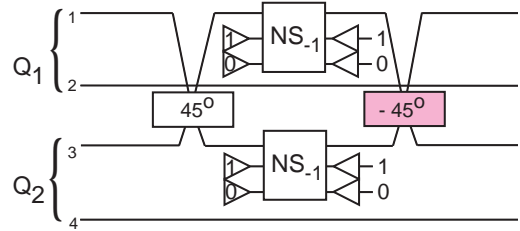}
\end{center}
\caption{\footnotesize Step 3 of a probabilistic CSign gate.}
\label{fig2KLMstep3Pic}
\end{figure}

This gate works with a probability of $\frac{1}{16}$. That is, there is
a probability of $\frac{1}{16}$ that we will measure the ancilla state
$\ket{10}$ in both $NS_{-1}$ gates.  It is important to notice that
when the result of the measurement of the ancilla state is other
than  $\ket{10}$, the gate fails but we know that this failure has occurred.

It is also interesting that during the gate,
between the $\pm 45^{\circ}$ beam splitters, the photons are not in 
the dual rail logic encoding but recombine at the last beam splitter 
to get back into the right encoding. Thus we go out of the qubits Hilbert space
to make this gate and go back at its output.

The state  
\begin{eqnarray}
 \frac{1}{2}\Bigl(\ket{0101}+\ket{0110}+\ket{1001}-\ket{1010}\Bigr)
\label{CSignStateEqn}
\end{eqnarray}
is the first building block to performing a CSign on photonic qubits
with probability arbitrarily close to 1. 

\subsubsection{Teleporting qubits through a gate}

What we have shown so far is how to perform a probabilistic CSign
gate on photonic qubits. This is not enough to allow scalable quantum
computation. The probabilistic CSign gate will be used as an entangled
state production stage.

To perform a two qubit gate we could teleport \cite{Bennet93} the
qubits through the gate as first shown by Gottesman and Chuang
\cite{Gottesman99}.

Say we have two arbitrary qubits $\ket{\psi_{1}}_{q}$ and $\ket{\psi_{2}}_{q}$ that
we want to perform a CSign on. We could teleport both these qubits and
then apply a CSign on them, as seen in the fig. below:

\begin{figure}[h!]
\begin{center}
\includegraphics[width=0.7\textwidth]{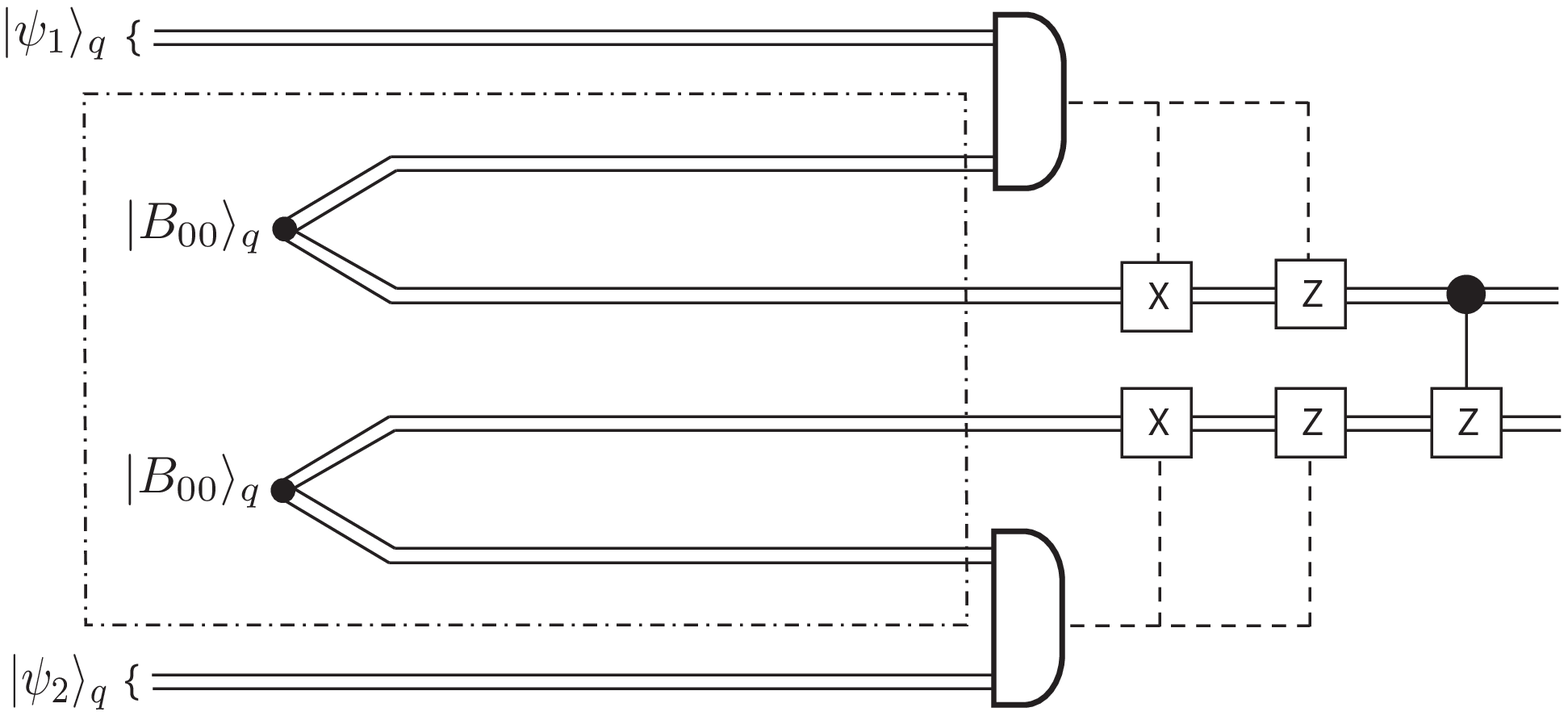}
\end{center}
\caption{\footnotesize An application a CSign between qubits  $\ket{\psi_{1}}_{q}$ and $\ket{\psi_{2}}_{q}$. The dashed-dotted box is the state preparation area. In this case just the Bell states $\ket{B_{00}}$ need to be prepared.}
\label{teleport1aPic}
\end{figure}

{\bf Exercise}:  given that $\ket{B_{00}}_{q}=\frac{1}{\sqrt{2}}\bigl(\ket{00}+\ket{11}\bigr)_{q}$,  $\ket{\psi_{1}}_{q}=c_{0}\ket{0}_{q}+c_{1}\ket{1}_{q}$, $\ket{\psi_{2}}_{q}=d_{0}\ket{0}_{q}+d_{1}\ket{1}_{q}$ (all logical qubits, $|c_{0}|^{2}+|c_{1}|^{2}=|d_{0}|^{2}+|d_{1}|^{2}=1$), and that we measure in the logical basis, show that this circuit gives a CSign between $\ket{\psi_{1}}_{q}$ and $\ket{\psi_{2}}_{q}$. 

\begin{figure}[h!]
\begin{center}
\includegraphics[width=0.7\textwidth]{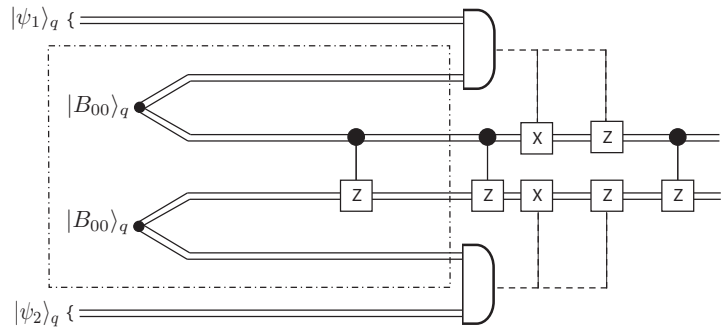}
\end{center}
\caption{\footnotesize An application a CSign between qubits $\ket{\psi_{1}}_{q}$ and $\ket{\psi_{2}}_{q}$.}
\label{teleport1bPic}
\end{figure}

Since CSign applied twice gives the identity, we can add two CSign's to fig. \ref{teleport1aPic} as seen in fig. \ref{teleport1bPic}. 
\newpage

We now have a CSign inside the dashed-dotted box, the state preparation area. Now we want to look at the effect of conjugating the ``corrections''  with the CSign's. To do this we will need the identity: 
\begin{eqnarray*}
\text{CSign}\Bigl(\sigma_{x}\otimes\id\Bigr)\text{CSign}&=& \sigma_{x}\otimes\sigma_{z}\\
\left( \begin{array}{cccc}
1&0&0&0\\
0&1&0&0\\
0&0&1&0\\
0&0&0&-1\\
\end{array} \right) 
\left( \begin{array}{cccc} 
0&0&1&0\\
0&0&0&1\\
1&0&0&0\\
0&1&0&0\\
\end{array} \right) 
\left( \begin{array}{cccc}
1&0&0&0\\
0&1&0&0\\
0&0&1&0\\
0&0&0&-1\\
\end{array} \right) 
&=&
\left( \begin{array}{cccc}
0&0&1&0\\
0&0&0&-1\\
1&0&0&0\\
0&-1&0&0\\
\end{array} \right) 
\end{eqnarray*}

\begin{figure}[h!]
\begin{center}
\includegraphics[width=1.0\textwidth]{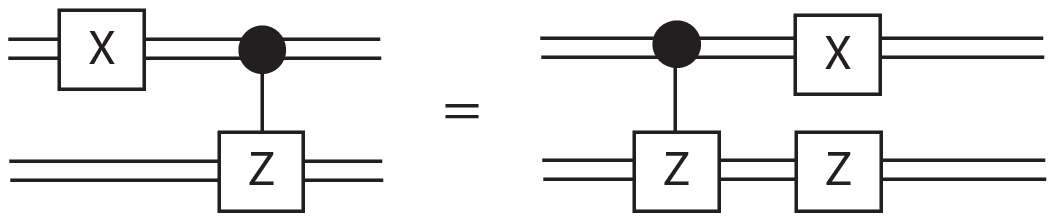}
\end{center}
\caption{\footnotesize Shows the identity 
$\text{CSign}\Bigl(\sigma_{x}\otimes\id\Bigr)\text{CSign}= 
\sigma_{x}\otimes\sigma_{z}$.}
\label{KLMboxdPic}
\end{figure}
and the identities
\begin{eqnarray*}
\text{CSign}\Bigl(\id\otimes \sigma_{x}\Bigr)\text{CSign}&=& 
\sigma_{z}\otimes\sigma_{x}\\
\text{CSign}\Bigl(\sigma_{z}\otimes\id\Bigr)\text{CSign}&=& 
\sigma_{z}\otimes\id\\
\text{CSign}\Bigl(\id\otimes \sigma_{z}\Bigr)\text{CSign}&=& 
\id\otimes\sigma_{z}
\end{eqnarray*}
Applying these identities gives the following circuit: 

\begin{figure}[h!]
\begin{center}
\includegraphics[width=0.7\textwidth]{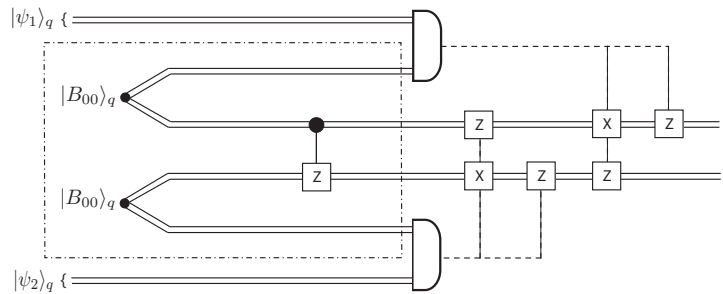}
\end{center}
\caption{\footnotesize An application a CSign between qubits $\ket{\psi_{1}}_{q}$ and $\ket{\psi_{2}}_{q}$.}
\label{teleportPic}
\end{figure}

The problem of performing a CSign has now been turned into a state
preparation problem. We can apply the CSign gate (fig. \ref{fig2KLMPic}) probabilistically offline. Once we know we have a
successful CSign state (eq. (\ref{CSignStateEqn})), we can proceed
and teleport the information qubits, as in fig.
\ref{teleportPic}. This way we do not corrupt the quantum
information.
 
\subsubsection{Teleporting with the C-Sign entangled states}

If a gate fails with probability $f$ and we perform error correction,
after we do $k$ of these gates we will have a success probability of
$f^{k}$. With teleportation, constructing two qubit gates reduces to a
state preparation problem. The critical element is that we know when
the state preparation fails and thus we can ensure that it does not
corrupt the quantum information. To succeed we need to attempt the
state preparation $\frac{k}{f}$ times to implement $k$ gates. 

The next step is to show how we perform teleportation with linear optics?

\subsubsection{Basic Teleportation with linear optics}

It is known~\cite{Lutkenhaus99} that we cannot distinguish between the four
Bell states with linear optics alone. With a single beam splitter the
best that we can do is distinguish between the Bell states, and thus do
teleportation,  with a
success probability of $\frac{1}{2}$. We will show how this is done
and then generalise the teleportation so that 
the probability of success is as near to one as desired. Although
we cannot achieve a probability of one as this would require perfect Bell state distinguishability, we can use quantum error correction and limit the left over error so that it is smaller than what is required from the accuracy threshold theorem. 

Consider teleporting the state $\alpha\ket{01}+\beta\ket{10}$.  We
need only look at teleporting the first mode of this state:

\begin{figure}[h!]
\begin{center}
\includegraphics[width=0.4\textwidth]{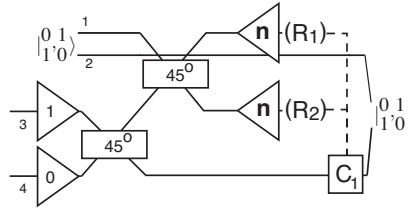}
\end{center}
\caption{\footnotesize Telelportation with linear optics.}
\label{sfig2KLMPic}
\end{figure}

First the entangled resource state $\frac{1}{\sqrt{2}}\bigl( \ket{01}_{34}+\ket{10}_{34}\bigr)$ is made, shown below. At this point the state is  
\begin{eqnarray*}
&\Bigl(&\alpha\ket{01}_{12}+\beta\ket{10}_{12}\Bigr)\frac{1}{\sqrt{2}}\Bigl( \ket{01}_{34}+\ket{10}_{34}\Bigr)\\
&=&\frac{1}{\sqrt{2}}\Bigl( \alpha\ket{0101}_{1234}+\alpha\ket{0110}_{1234} +\beta\ket{1001}_{1234}+\beta\ket{1010}_{1234}\Bigr)
\end{eqnarray*}

\begin{figure}[h!]
\begin{center}
\includegraphics[width=0.4\textwidth]{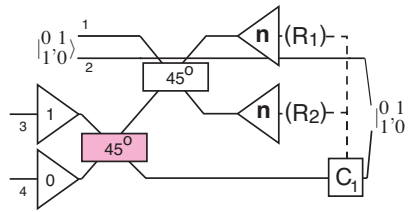}
\end{center}
\caption{\footnotesize Telelportation with linear optics:  making the entanglement resource state.}
\label{sfig2KLMstep1Pic}
\end{figure}

\newpage

After the second beam splitter we have:
\begin{eqnarray*}
\to \frac{1}{2}\Bigl(\sqrt{2}\alpha\ket{0101}-\alpha\ket{1100} +\alpha\ket{0110}+\beta\ket{1001}+\beta\ket{0011}-\beta\ket{2000}+\beta\ket{0020}\Bigr)
\end{eqnarray*}
Remembering that the second mode was unchanged here. 
\begin{figure}[h!]
\begin{center}
\includegraphics[width=0.4\textwidth]{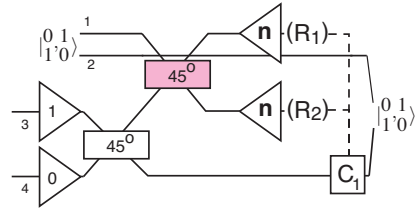}
\end{center}
\caption{\footnotesize  Telelportation with linear optics:  performing a partial Bell measurement.}
\label{sfig2KLMstep2Pic}
\end{figure}

If the detectors measure a total of 1 photon ($R_{1}+R_{2}=1$), we can recover the original information in modes 4 and 2:
\begin{eqnarray*}
R_{1}=0, R_{2}=1&:&  \quad \frac{1}{2}\bigl( \alpha\ket{01}+\beta\ket{10}\bigr)\\
R_{1}=1, R_{2}=0&:&  \quad \frac{1}{2}\bigl(- \alpha\ket{01}+\beta\ket{10}\bigr)\\
\end{eqnarray*}

\begin{figure}[h!]
\begin{center}
\includegraphics[width=0.4\textwidth]{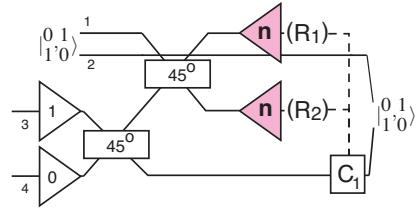}
\end{center}
\caption{\footnotesize  Telelportation with linear optics:  photo-detection.}
\label{sfig2KLMstep3Pic}
\end{figure}

\newpage 

In the case that we measure $R_{1}=1, R_{2}=0$ we need to
perform a phase shift correction gate $C_{1}$ and we have seen how to
do that in the previous section. The probability of successful
teleportation is $\frac{1}{4}+\frac{1}{4}=\frac{1}{2}$. If the detectors measure 0 or 2 photons the teleportation fails  by collapsing the state in the mode $\ket{0}$ or $\ket{1}$ and this is equivalent to projecting the qubit in the $Z$ basis, as will be explained in the next section. This occurs with a probability of $\frac{1}{2}$.

\begin{figure}[h!]
\begin{center}
\includegraphics[width=0.4\textwidth]{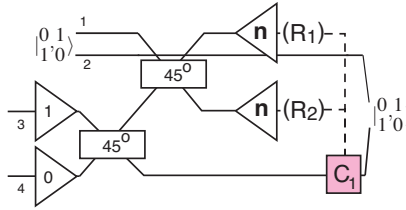}
\end{center}
\caption{\footnotesize Telelportation with linear optics:  applying a correction.}
\label{sfig2KLMstep4Pic}
\end{figure}

\subsubsection{The teleported C-Sign}

Now we can piece together what we have worked out so far to make a CSign
gate. First we have the CSign entangled state production step:

\begin{figure}[h!]
\begin{center}
\includegraphics[width=0.7\textwidth]{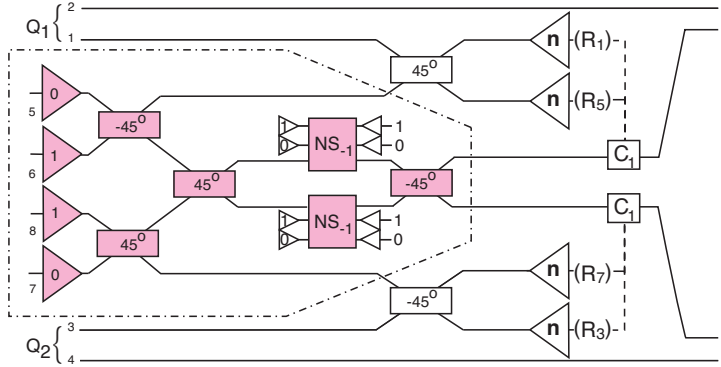}
\end{center}
\caption{\footnotesize CSign state production}
\label{fig3KLMPic3}
\end{figure}

\newpage
Followed by the teleportation: 

\begin{figure}[h!]
\begin{center}
\includegraphics[width=0.7\textwidth]{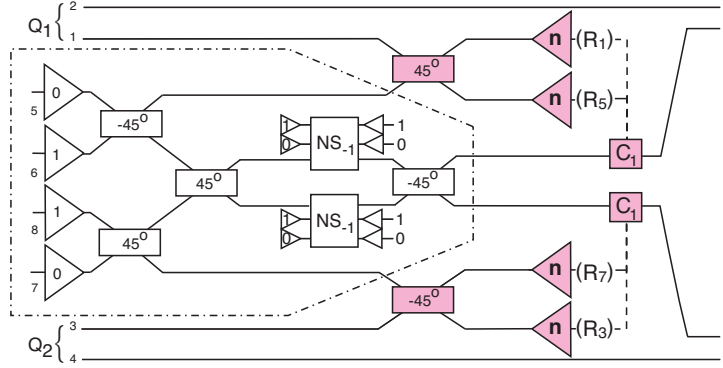}
\end{center}
\caption{\footnotesize Teleportation.}
\label{fig3KLMPic4}
\end{figure}

The success probability for the state production stage (fig. \ref{fig3KLMPic3}) is $\frac{1}{16}$, however this is done
offline. This state preparation fails (on average 16 times) 
when the measurement of the ancilla give the wrong result and then we just need to repeat the state preparation. This failure does not corrupt the quantum information as the qubtis have not yet interacted with these photons.  When the measurement of the ancilla give the right values, we know that the state is 
the correct one for teleportation. 

The success probability for this CSign gate is
$\frac{1}{2}$. 
{\bf Exercise}: given that the dashed-dotted box in fig.
\ref{fig3KLMPic3} makes a state of the form $\frac{1}{2}\bigl(
\ket{1010}+\ket{1001}-\ket{0110}+\ket{0101}\bigr)_{5687}$, show that
this circuit executes a C-sign between $Q_{1}$ and $Q_{2}$. 
{\bf Exercise}: what error occurs on the information qubits if the
detectors $R_{1},R_{3}, R_{5}$ and $R_{7}$ do not measure the correct sequence?

\subsubsection{Increasing the probability of success}

The task that remains now is to improve the probability of success
of the teleportation.
If we increase the complexity of the state produced in the dashed-dotted box
of fig. \ref{fig3KLMPic3}, we can increase the teleportation success
probability asymptotically close to one. We can generalise eq.
(\ref{CSignStateEqn}) by increasing the number of modes:
\begin{eqnarray*}
\ket{t_{n}} = 
\frac{1}{\sqrt{n+1}}\sum_{j=0}^{n} \ket{1}^{j}\ket{0}^{n-j}\ket{0}\ket{1}^{n-j}
\end{eqnarray*}
where the exponent $k$ means that this state is repeated $k$ times, i.e.
$\ket{1}^2\rightarrow \ket{1}\ket{1}$.

Or in the logical basis 
\begin{eqnarray*}
 \ket{t_{n}}_q = \frac{1}{\sqrt{n+1}}\sum_{j=0}^{n} \ket{0}_q^{j}\ket{1}_q^{n-j}
\end{eqnarray*}
When we use this preparation state for teleportation the success
probability scales as $\frac{n}{n+1}=1-\frac{1}{n+1}$.

\subsubsection{Generalised beam splitter} 

When we use the larger entangled state $\ket{t_{n}}$ we need to use a generalised beam splitter:  the $n+1$ point Fourier transform. 

We define the $n+1$ point Fourier transform with $\hat{F}_{n+1}$. The matrix elements are given by 
\begin{eqnarray*}
u\bigl(\hat{F}_{n+1}\bigr)_{kl}=\frac{e^{i \frac{2\pi k l}{n+1}}}{\sqrt{n+1}}
\end{eqnarray*}
{\bf Exercise}:  show that for $n=1$ we get back our anti-symmetric 50-50 beam splitter transformation.

As in fig. \ref{sfig2KLMstep3Pic}, we measure all the outputs of the $n+1$ point Fourier transform. If we detect $k$ photons and $0<k<n+1$, then we know the qubit has been teleported to mode $n+k$. If we measure either 0 or $n+1$ photons the teleportation has failed. 

For example, consider the case when $n=2$: 
\begin{eqnarray*}
\ket{t_{2}}=\frac{1}{\sqrt{3}}\Bigl( \ket{0011}+\ket{1001}+\ket{1100}\Bigr)
\end{eqnarray*}
and
\begin{eqnarray*}
u(F_{3})=\left( \begin{array}{ccc}
1&1&1\\
1&e^{\frac{2\pi i}{3}}&e^{-\frac{2\pi i}{3}}\\
1&e^{-\frac{2\pi i}{3}}&e^{\frac{2\pi i}{3}}\\
\end{array}\right)
\end{eqnarray*}

{\bf Exercise:} Find the linear circuit that implement this unitary
transformation.

\begin{figure}[h!]
\begin{center}
\includegraphics[width=1.0\textwidth]{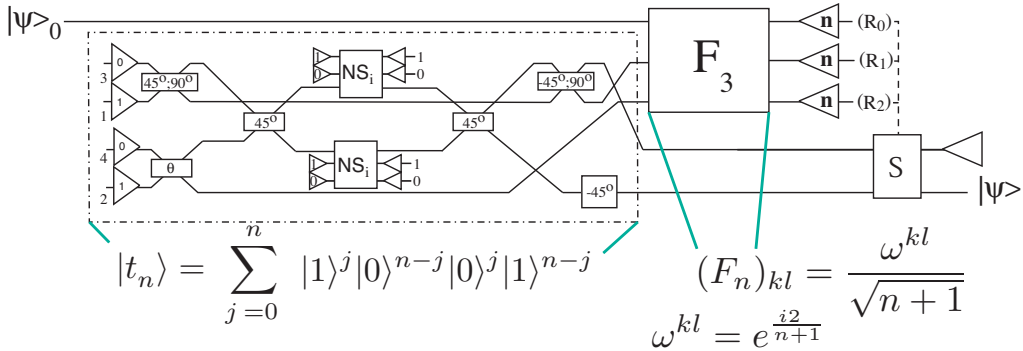}
\end{center}
\caption{\footnotesize Teleportation with $\ket{t_{2}}$. The dashed-dotted box produces $\ket{t_{2}}$. The probability of success of this teleportation is 2/3.}
\label{gteleportPic}
\end{figure}

When $F_{3}$ is applied to the incident modes 0, 1 and 2, the modes are transformed as follows: 
\begin{eqnarray*}
\hat{a}_{0}^{\dagger}&\to& \frac{1}{\sqrt{3}}\bigl( \hat{a}_{0}^{\dagger}+\hat{a}_{1}^{\dagger}+\hat{a}_{2}^{\dagger}\bigr)\\
\hat{a}_{1}^{\dagger}&\to& \frac{1}{\sqrt{3}}\bigl( \hat{a}_{0}^{\dagger}+e^{\frac{2\pi i}{3}}\hat{a}_{1}^{\dagger}+e^{-\frac{2\pi i}{3}}\hat{a}_{2}^{\dagger}\bigr)\\
\hat{a}_{2}^{\dagger}&\to& \frac{1}{\sqrt{3}}\bigl( \hat{a}_{0}^{\dagger}+e^{-\frac{2\pi i}{3}}\hat{a}_{1}^{\dagger}+e^{\frac{2\pi i}{3}}\hat{a}_{2}^{\dagger}\bigr)\\
\end{eqnarray*}
Now say we want to teleport the state $\alpha\ket{01}+\beta\ket{10}$
(where the second mode of this state propagates unchanged) with the
entangled resource state $\ket{t_{2}}$.

Once we apply the 3 point Fourier transform to the states
$\alpha\ket{01}+\beta\ket{10}$ and $\ket{t_{2}}$, we have 19
measurement outcomes: $R_{0},R_{1},R_{2}=$ 000, 001, 010, 100, 011,
101, 110, 002, 020, 200, 012, 021, 201, 102, 120, 210, 300, 030 and
003. We are only interested in measuring total photon numbers between
0 and $n+1=3$, that is, only 1 or 2 photons in total.

The measurement outcomes we are interested in are given in the table
below, along with the mode the information is output in and the
probability of that measurement. The following identities will be
useful when working through this example: $e^{i\frac{2}{3}\pi
}+e^{-i\frac{2}{3}\pi }=-1$, $e^{i\frac{4}{3}\pi}+e^{-i\frac{4}{3}\pi
}=-1$, $e^{i\frac{2}{3}\pi }+1=e^{i \frac{\pi}{3}}$ and
$e^{-i\frac{2}{3}\pi }+1=e^{ -i\frac{\pi}{3}}$

The total probability for successful teleportation is
$3\bigl(\frac{1}{9}\bigr)+3\bigl(\frac{2}{27}\bigr)+3\bigl(\frac{1}{27}\bigr)
=\frac{2}{3}$. This corresponds to $\frac{n}{n+1}$ with n=2. As in the
case of the previous subsubsection, when the teleportation fails we know
it and then the state is projected in the $Z$ basis.

\begin{table}[h]
\begin{tabular}{cccc}
\hline
Measurement & Output&Mode&Probability\\ 
\hline 
$\ket{100}$&$\frac{1}{3}\bigl(\alpha\ket{0}+\beta\ket{1}\bigr)$&3&$\frac{1}{9}$\\
$\ket{010}$&$\frac{1}{3}\bigl(e^{\frac{2\pi i}{3}}\alpha\ket{0}+\beta\ket{1}\bigr)$&3&$\frac{1}{9}$\\
$\ket{001}$&$\frac{1}{3}\bigl(e^{-\frac{2\pi i}{3}}\alpha\ket{0}+\beta\ket{1}\bigr)$&3&$\frac{1}{9}$\\
$\ket{200}$&$\frac{\sqrt{2}}{3\sqrt{3}}\bigl(\alpha\ket{0}+\beta\ket{1}\bigr)$&4&$\frac{2}{27}$\\
$\ket{020}$&$\frac{\sqrt{2}}{3\sqrt{3}}\bigl(\alpha\ket{0}+e^{\frac{2\pi i}{3}}\beta\ket{1}\bigr)$&4&$\frac{2}{27}$\\
$\ket{002}$&$\frac{\sqrt{2}}{3\sqrt{3}}\bigl(\alpha\ket{0}+e^{-\frac{2\pi i}{3}}\beta\ket{1}\bigr)$&4&$\frac{2}{27}$\\
$\ket{110}$&$\frac{1}{3\sqrt{3}}\bigl(-\alpha\ket{0}+e^{\frac{i\pi }{3}}\beta\ket{1}\bigr)$&4&$\frac{1}{27}$\\
$\ket{101}$&$\frac{1}{3\sqrt{3}}\bigl(-\alpha\ket{0}+e^{-\frac{i\pi }{3}}\beta\ket{1}\bigr)$&4&$\frac{1}{27}$\\
$\ket{011}$&$\frac{1}{3\sqrt{3}}\bigl(-\alpha\ket{0}-\beta\ket{1}\bigr)$&4&$\frac{1}{27}$\\
\end{tabular}
\caption{\footnotesize The telelported states when using $\ket{t_{2}}$ as an entanglement resource.}
\end{table} 

\subsubsection{Bounds on success probabilities}

We have shown in this lecture that we can perform the $NS_{-1}$ gate
operation $\alpha\ket{0}+\beta\ket{1}+\gamma\ket{2}\to
\alpha\ket{0}+\beta\ket{1}-\gamma\ket{2}$ with a success probability
of $\frac{1}{4}$ (fig. \ref{fig1KLMPic}). We have also shown that
using two of these $NS_{-1}$ gates we can perform a CSign with success
probability $\frac{1}{16}$ (fig. \ref{fig2KLMPic}). However, it is
not known whether these probabilities are optimal.

\begin{figure}[h!]
\begin{center}
\includegraphics[width=0.9\textwidth]{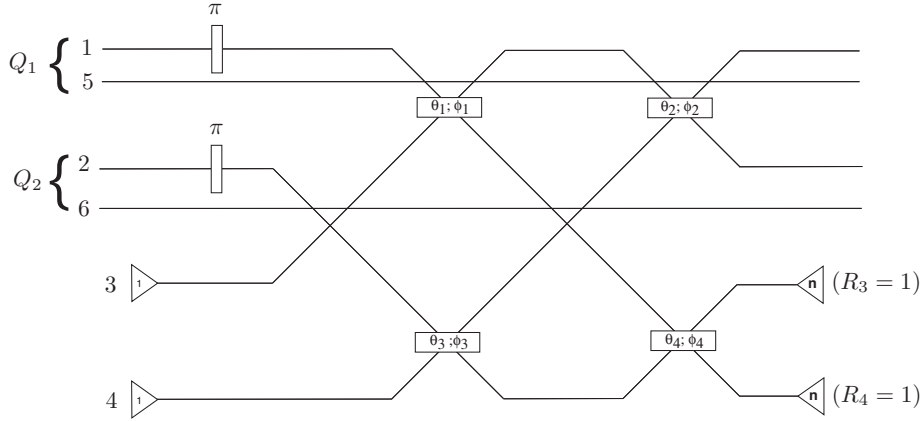}
\end{center}
\caption{\footnotesize A CSign gate with success probability $\frac{2}{27}$. Here $\theta_{1}=54.74^{o}, \phi_{1}=0$, $\theta_{2}=-54.74^{o}, \phi_{2}=0$, $\theta_{3}=54.74^{o}, \phi_{3}=0$ and $\theta_{4}=17.63^{o}, \phi_{4}=0$.}
\label{Knillcs2_27Pic}
\end{figure}

To date, the best $NS_{-1}$ gate success probability is as above and
the best CSign gate succeeds with a probability of $\frac{2}{27}$
\cite{Knill02}. The $\frac{2}{27}$ CSign gate is shown in fig. \ref{Knillcs2_27Pic}.

In this circuit the two qubits 
$Q_{1}:  \alpha\ket{01}_{15}+\beta\ket{10}_{15}$ and
$Q_{2}:  \gamma\ket{01}_{26}+\delta\ket{10}_{26}$ are to have a CSign applied,
where modes 5 and 6 are left unchanged. {\bf   Exercise}: check that this circuit implements a CSign on measuring single photons in modes 3 and 4.

Knill has shown that the CSign gate has an upper bound of
$\frac{3}{4}$ when using two ancilla modes, each with a maximum of 1
photon per mode \cite{Knill03}. Knill has also shown that the upper
bound for any $NS_{-1}$ gate with one ancilla mode is $\frac{1}{2}$
\cite{Knill03}. Scheel \textit{et. al.} conjecture that the upper
bound for the $NS_{-1}$ gate is $\frac{1}{4}$, regardless of the
dimensionality of the ancilla space \cite{Scheel04}.


\section{LOQC and Quantum Error Correction}

\subsection{Improving LOQC:  beyond state preparation}\label{BeyondStatePrepSect}

In the last section we showed that scalable quantum computing was
possible with linear optical elements, single photon sources and
photo-detection. The problem of making a two qubit gate with
probability of success arbitrarily close to one was transferred to a
state preparation problem. By using generalised entangled states of
the form $\ket{t_{n}}$, the probability of success scaled as
$\frac{n}{n+1}$. However, states of the form $\ket{t_{n}}$ are
complicated to make. To date the best schemes need resources that are
exponential in $n$ \cite{Myers02, Franson04}. Instead, we may ask
ourselves if we can incorporate quantum error correction
\cite{Knill02a} into the LOQC proposal. Is it possible to correct for
the incorrect measurements in the basic linear optical teleportation
(fig. \ref{sfig2KLMPic})? Can we use quantum error correction along
with smaller $\ket{t_{n}}$ states to increase the probability of
successful gates giving scalable quantum computing?

To answer this we first need to know what errors are inflicted on the
quantum information when an incorrect measurement is made in the
teleportation in fig. \ref{sfig2KLMPic}.

In the previous section we went through this basic teleportation with
linear optics and showed that if we want to teleport the state
$\alpha\ket{01}+\beta\ket{10}$ using the entanglement resource state
$\frac{1}{\sqrt{2}}\bigl( \ket{01}+\ket{10}\bigr)$, we need to measure
1 and only 1 photon at the photo-detection stage (fig. \ref{sfig2KLMstep3Pic}). The probability of success was
$\frac{1}{2}$. But what happens when we measure 0 or 2 photons?

Recalling from the last section,  our input state is transformed to 
\begin{eqnarray*}
\to \frac{1}{2}\Bigl( \sqrt{2}\alpha\ket{0101}-\alpha\ket{1100}
+\alpha\ket{0110}+\beta\ket{1001}+\beta\ket{0011}-\beta\ket{2000}
+\beta\ket{0020}\Bigr)
\end{eqnarray*}
just before the measurement in fig. \ref{sfig2KLMstep3Pic}
(remembering the 2nd mode is untouched). If the detectors measure a
total of 1 photon ($R_{1}+R_{2}=1$) we can recover the original
information:
\begin{eqnarray*}
R_{1}=0, R_{2}=1&:&  \quad 
\frac{1}{2}\bigl( \alpha\ket{01}+\beta\ket{10}\bigr)\\
R_{0}=1, R_{1}=1&:&  \quad 
\frac{1}{2}\bigl(- \alpha_{0}\ket{01}+\alpha_{1}\ket{10}\bigr)
\end{eqnarray*}

If the detectors measure a total of 0 or 2 photons ($R_{1}+R_{2}=0\,\,
\text{mod}\,\,2$) we have:
\begin{eqnarray*}
R_{1}=0, R_{2}=0&:&  \quad \frac{1}{\sqrt{2}}\alpha_{0}\ket{11}\\
R_{0}=2, R_{1}=0&:&  \quad -\frac{1}{\sqrt{2}}\alpha_{1}\ket{00}\\
R_{1}=0, R_{2}=2&:&  \quad \frac{1}{\sqrt{2}} \alpha_{1}\ket{00}\\
\end{eqnarray*}
Note that when we measure 0 or 2 photons we have measured our qubit,
collapsing our ``quantum'' information. This means we are free to add
a single photon, or vacuum mode. When we measure 0 photons, we obtain
the state $\ket{01}=\ket{0}_{q}$ and when we measure 2 photons, we
obtain $\ket{10}=\ket{1}_{q}$. When we measure 0 or 2 photons we
effectively measure our information qubit in the $Z$-basis. 
{\bf Exercise}: check that the probability of measuring 0, 1 or 2 photons
in total adds to 1.

Knowing this we should be able to use quantum error correction to
somehow encoded our quantum data and then correct when the
teleportation fails, i.e. when the measurement give either 0 or 2
photons in fig. \ref{sfig2KLMstep3Pic}. This means we need to find
quantum error correcting codes that deal with an error model corresponding
with  projection in the $Z$ basis, where we know which qubit was
projected.  From this point on we'll be dealing with Z-measurement errors
that are appearing when we try to make a two bit gate on the qubits.
We will assume that the one bit gates (implemented by linear optics)
are error free. But first we need to know what a quantum error correcting code is. 

\subsection{Quantum Error Correcting Codes}

\subsubsection{What are they? }

We can encode our quantum information with code words to protect
against errors. Classically, when the error model is given by independent
bit flips  $(X)$,  this can be done via the repetition code:  $0\to 000$ and $1\to 111$.

When we generalise the classical theory to the quantum one,
we also have to worry about phase errors ($Z$ operators 
and the combination of bit flip and phase errors:  $Y$ operators). A
necessary and sufficient condition for a code with basis code words
$\{ \ket{\psi}\}$ to correct for errors $\{ E\}$ is
\begin{eqnarray}
\bra{\psi_{i}}E_{a}^{\dagger}E_{b}\ket{\psi_{j}}=C_{ab}\delta_{ij}\label{QECCcond}
\end{eqnarray}
where $\ket{\psi_{i}},\ket{\psi_{j}}\in\{ \ket{\psi}\}$ and 
$E_{a}, E_{b}\in\{ E\}$.

Say we have errors of the form 
$E_{1}=\frac{1}{2}\bigl( \id +Z\bigr)$, 
$E_{2}=\frac{1}{2}\bigl( \id - Z\bigr)$. What does this do
to our qubit $\ket{\psi}=\alpha\ket{0}_{q}+\beta\ket{1}_{q}$?
\begin{eqnarray*}
E_{1}\ket{\psi}&=&\alpha\ket{0}_{q}\\
E_{2}\ket{\psi}&=&\beta\ket{1}_{q}
\end{eqnarray*}
We have measured our qubit in the $Z$-basis. How can we encode to
correct for this error? We could try the following encoding:
\begin{eqnarray}
\ket{\overline{0}}&=&\frac{1}{\sqrt{2}}\bigl(\ket{00}_{q}+\ket{11}_{q}\bigr)
\label{Zmeas0Eqn}\\ 
\ket{\overline{1}}&=&\frac{1}{\sqrt{2}}\bigl(\ket{01}_{q}+\ket{10}_{q}\bigr)
\label{Zmeas1Eqn}
\end{eqnarray} 

If we measure the first mode of our encoded qubit in the $Z$-basis we have
\begin{eqnarray*}
\frac{1}{\sqrt{2}}\bigl(&\alpha&
\ket{00}_{q}+\alpha\ket{11}_{q}+\beta\ket{01}_{q}+\beta\ket{10}_{q}\bigr)\\
&=&\frac{1}{\sqrt{2}}
\{\ket{0}_{q}\bigl(\alpha\ket{0}_{q}+\beta\ket{1}_{q}\bigr)\}
+\frac{1}{\sqrt{2}} 
\ket{1}_{q}\bigl(\alpha\ket{1}_{q}+\beta\ket{0}_{q}\bigr)\}
\end{eqnarray*}
and we see that if we measure mode 1 to be a $\ket{0}_{q}$, we get our
original information qubit back and if we measure mode 1 to be a
$\ket{1}_{q}$, we can get our original information qubit back once we
apply a bit flip. Now we need to check that this code satisfies
condition \ref{QECCcond} for the errors $E_{1}$ and $E_{2}$. We assume
the errors are on the 1st qubit, so $E_{1}=\frac{1}{2}\bigl( \id +
Z\bigr)\otimes \id$ and $E_{2}=\frac{1}{2}\bigl( \id -
Z\bigr)\otimes\id$:
\begin{table}[h!]
\begin{tabular}{cccc}
$\bra{\overline{0}}E_{1}^{\dagger}\id\ket{\overline{0}}=\frac{1}{2}$
&$\bra{\overline{0}}E_{2}^{\dagger}\id\ket{\overline{0}}=\frac{1}{2}$
&$\bra{\overline{0}}E_{1}^{\dagger}E_{2}\ket{\overline{0}}=0$
&$\bra{\overline{0}}E_{2}^{\dagger}E_{1}\ket{\overline{0}}=0$\\
$\bra{\overline{0}}E_{1}^{\dagger}\id\ket{\overline{1}}=0$
&$\bra{\overline{0}}E_{2}^{\dagger}\id\ket{\overline{1}}$=0
&$\bra{\overline{0}}E_{1}^{\dagger}E_{2}\ket{\overline{1}}=0$
&$\bra{\overline{0}}E_{2}^{\dagger}E_{1}\ket{\overline{1}}=0$\\
$\bra{\overline{1}}E_{1}^{\dagger}\id\ket{\overline{0}}=0$
&$\bra{\overline{1}}E_{2}^{\dagger}\id\ket{\overline{0}}$=0
&$\bra{\overline{1}}E_{1}^{\dagger}E_{2}\ket{\overline{0}}=0$
&$\bra{\overline{1}}E_{2}^{\dagger}E_{1}\ket{\overline{0}}=0$\\
$\bra{\overline{1}}E_{1}^{\dagger}\id\ket{\overline{1}}=\frac{1}{2}$
&$\bra{\overline{1}}E_{2}^{\dagger}\id\ket{\overline{1}}=\frac{1}{2}$
&$\bra{\overline{1}}E_{1}^{\dagger}E_{2}\ket{\overline{1}}=0$
&$\bra{\overline{1}}E_{2}^{\dagger}E_{1}\ket{\overline{1}}=0$\\
\end{tabular}
\caption{\footnotesize Testing condition \ref{QECCcond} for the errors 
$E_{1}$ and $E_{2}$.}
\end{table} 
\newpage
But this doesn't tell us how to correct for the errors $E_{1}$ and
$E_{2}$, only that correction is possible.

\subsubsection{$Z$-measurement Quantum Error Correcting Code (QECC)}
 
How can we correct for the Z-measurement errors $E_{1}$ and $E_{2}$?
There is a formalism called the stabilizer formalism (see
\cite{Gottesman97}) that gives us a way to do this. The stabilizer is
defined by a set of operators for an $n$ qubit system whose common
eigenvectors define a $2^{k}$-dimensional subspace (the code). That
is, the stabilizer is defined as the set of operators $\{M_{i}\}$ that
leave the code word space $\{\ket{\overline{\psi_{j}}}\}$ invariant:
$M_{i}\ket{\overline{\psi_{j}}}=\ket{\overline{\psi_{j}}}\,\,\forall
i,j$. We will focus on operators formed by tensor products of Pauli
operators. We have the two code words:
$\ket{\overline{0}}=\frac{1}{\sqrt{2}}\bigl(\ket{00}_{q}+\ket{11}_{q}\bigr)$
and
$\ket{\overline{1}}=\frac{1}{\sqrt{2}}\bigl(\ket{01}_{q}+\ket{10}_{q}\bigr)$.
These are stabilized by the two operators $\id\otimes\id$ and
$X\otimes X$. 
{\bf Exercise}: check that $XX$ stabilizer the
$Z$-measurement code.

For a stabilizer code we can detect all errors that are either in the
stabilizer or anti-commute with any member of the stabilizer
group. Say we have the QECC with code words
$\ket{\overline{\psi}_{j}}$ and $M_{i}$ is an element of the
stabilizer $S$. And say $E$ is an error such that $\{M_{i},E\}=0$,
then
$M_{i}E\ket{\overline{\psi}_{j}}=
-EM_{i}\ket{\overline{\psi}_{j}}=-E\ket{\overline{\psi}_{j}}$. So
$E\ket{\overline{\psi}_{j}}$ is an eigenstate of $M_{i}$, so measuring
$M_{i}$ will tell us if $E$ has occurred. By measuring all the
stabilizer generators we identify the error syndrome with this method.

If a QECC can correct for errors $E$ and $F$, then it can also correct
for the error $aE+bF$. From the form of $E_{1}$ and $E_{2}$ we need
only consider correcting the error $Z\otimes \id$. More generally,
when we consider correcting for a $Z$-measurement error on either
qubit of the code words (\ref{Zmeas0Eqn}), (\ref{Zmeas1Eqn}), we need to
consider the errors $Z\otimes \id$ and $\id\otimes Z$. Both $Z\otimes
\id$ and $\id\otimes Z$ anti-commute with $X\otimes X$. {\bf
  Exercise}: check this.

The operator $X\otimes X$ has the matrix form:
\begin{eqnarray*}
X\otimes X=\left( 
  \begin{array}{cc}  0 & 1  \\
                     1 & 0  \\ 
\end{array} \right) 
\otimes\left( 
  \begin{array}{cc}  0 & 1  \\
                     1 & 0  \\ 
\end{array} \right) =
\left( 
\begin{array}{cccc}  0&0&0&1  \\ 
                     0&0&1&0  \\ 
                     0&1&0&0  \\ 
                     1&0&0&0  \\ 
\end{array} \right)   
\end{eqnarray*}
and has eigenvectors:   
\begin{eqnarray*}
\ket{XX_{10,00}}&=&\bigl( \pm\ket{00}_{q}+\ket{11}_{q}\bigr)/\sqrt{2}\\
\ket{XX_{11,01}}&=&\bigl( \pm\ket{01}_{q}+\ket{10}_{q}\bigr)/\sqrt{2}
\end{eqnarray*}
with the following eigenvalues:  
\begin{eqnarray*}
X\otimes X\ket{XX_{00}}&=&-\ket{XX_{00}}\\
X\otimes X\ket{XX_{01}}&=&-\ket{XX_{01}}\\
X\otimes X\ket{XX_{10}}&=&+\ket{XX_{10}}\\
X\otimes X\ket{XX_{11}}&=&+\ket{XX_{11}}
\end{eqnarray*}

For example, consider the first qubit of our encoded state
$\ket{\overline{\psi}}=\alpha\ket{\overline{0}}+\beta\ket{\overline{1}}
=\frac{1}{\sqrt{2}}\bigl(\alpha\ket{00}_{q}+\alpha\ket{11}_{q}
                        +\beta\ket{01}_{q}+\beta\ket{10}_{q}\bigr)$
is measured in the $Z$-basis:
\begin{eqnarray*}
\ket{\overline{\psi}}\to\frac{1}{\sqrt{2}}
\ket{0}_{q}\bigl( \alpha\ket{0}_{q}+\beta\ket{1}_{q}\bigr) 
\quad\textrm{or}\quad
\frac{1}{\sqrt{2}}
\ket{1}_{q}\bigl( \alpha\ket{1}_{q}+\beta\ket{0}_{q}\bigr)
\end{eqnarray*}
What if we measure the first qubit in the $+Z$ eigenstate 
$\ket{0}_{q}$? $\ket{\overline{\psi}}$ becomes:
\begin{eqnarray*}
&&\frac{1}{2}\bigl( \alpha\ket{XX_{10}}-\alpha\ket{XX_{00}}
                    +\beta\ket{XX_{11}}-\beta\ket{XX_{01}}\bigr)
\end{eqnarray*}

If we measure $X\otimes X$ and get $+1$ we have:
$\frac{1}{2\sqrt{2}}\bigl(\alpha\ket{00}_{q}+\alpha\ket{11}_{q}
                         +\beta\ket{01}_{q}+\beta\ket{10}_{q}\bigr)$,
our original encoded state. If we measure $X\otimes X$ and get $-1$ we
have:
$\frac{1}{2\sqrt{2}}\bigl(\alpha\ket{00}_{q}-\alpha\ket{11}_{q}
                          +\beta\ket{01}_{q}-\beta\ket{10}_{q}\bigr)$,
and we need to perform the $Z$ operation on the first qubit to get
back our original encoded state.

\begin{figure}[h!]
\begin{center}
\includegraphics[width=0.7\textwidth]{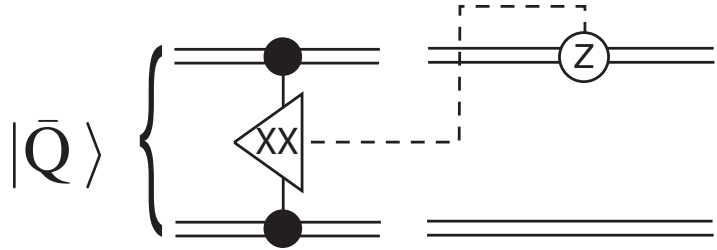}
\end{center}
\caption{\footnotesize Quantum error correction for $Z$-measurement errors.}
\label{projxx1Pic}
\end{figure}

Now what if we measure the first qubit in the $-Z$ eigenstate
$\ket{1}_{q}$? $\ket{\overline{\psi}}$ becomes:
\begin{eqnarray*}
&&\frac{1}{2}\bigl( \alpha\ket{XX_{00}}+\alpha\ket{XX_{10}}
                    +\beta\ket{XX_{01}}+\beta\ket{XX_{11}}\bigr)
\end{eqnarray*}
If we measure $X\otimes X$ and get $+1$ we have:
$\frac{1}{2\sqrt{2}}\bigl(\alpha\ket{00}_{q}+\alpha\ket{11}_{q}
                          +\beta\ket{01}_{q}+\beta\ket{10}_{q}\bigr)$,
our original encoded state. If we measure $X\otimes X$ and get $-1$ we
have:
$\frac{1}{2\sqrt{2}}\bigl(-\alpha\ket{00}_{q}+\alpha\ket{11}_{q}
                          -\beta\ket{01}_{q}+\beta\ket{10}_{q}\bigr)$,
and we need to perform the $Z$ operation on the first qubit to get
back our original encoded state (leaving us with an overall phase
factor). This can be summarised in fig. \ref{projxx1Pic}.

\begin{figure}[h!]
\begin{center}
\includegraphics[width=0.7\textwidth]{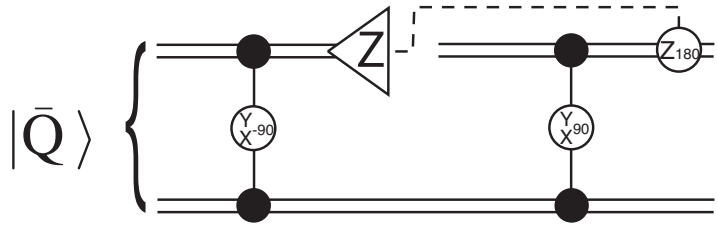}
\end{center}
\caption{\footnotesize An alternate quantum error correction circuit for $Z$-measurement errors.}
\label{projxx2Pic}
\end{figure}

Alternatively, a simpler method is shown in fig. \ref{projxx2Pic}, where we have 
$(UV)_{\theta}=(U\otimes V)_{\theta}
=\cos\frac{\theta}{2}-i\sin\frac{\theta}{2}\,\,(UV)$ and 
$(U\otimes V)^{2}=\id$.
{\bf  Exercise}: show that fig. \ref{projxx2Pic} is equivalent to fig. \ref{projxx1Pic}, correcting for $Z$-measurement errors on the top qubit.
\newpage
\subsection{Properties of the $Z$-measurement QECC}

\subsubsection{State Preparation}

\begin{figure}[h!]
\begin{center}
\includegraphics[width=0.7\textwidth]{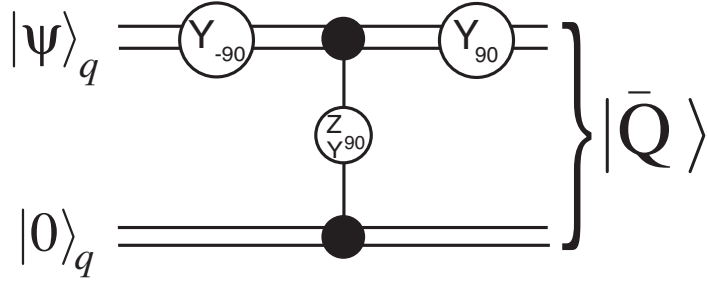}
\end{center}
\caption{\footnotesize State preparation for the $Z$-measurement code. }
\label{KLMboxgPic}
\end{figure}

If we start with the state
$\ket{\psi}=\alpha\ket{0}_q+\beta\ket{1}_q$, how can we prepare the
state $\ket{\bar\psi}=\alpha\ket{\bar 0}+\beta\ket{\bar1}$? To do this
we need an ancilla $\ket{0}_{q}$ state, two single qubit rotations and
one two qubit rotation, as seen in fig. \ref{KLMboxgPic}.

 {\bf Exercise}: what single qubit rotations are needed to
make $ZZ_{90}$ from a $ZY_{90}$? Note that depending on the choice of
$\alpha$ and $\beta$ we can prepare the encoded eigenstates of $X$,
$Y$ and $Z$. $X$, $Y$ and $Z$ eigenstate production are error
free. 
{\bf Exercise}: check that fig. \ref{KLMboxgPic} performs the
desired transformation on $\alpha\ket{0}_{q}+\beta\ket{1}_{q}$.
  
\subsubsection{Single qubit rotations}

How do we perform encoded $X$, $Y$ and $Z$ operations? It is to verify
that the following operators indeed perform the desired operations:

\begin{eqnarray*}
\overline{X}&=&X^{(1)}\,\,(\text{or}\,\, X^{(2)})\\
\overline{Y}&=&Y^{(1)}Z^{(2)}\,\,(\text{or}\,\, Z^{(1)}Y^{(2)})\\
\overline{Z}&=&Z^{(1)}Z^{(2)}\,\,(\text{or}\,\, -Y^{(1)}Y^{(2)})
\end{eqnarray*}
{\bf Exercise}:  check this. 

Any single qubit rotation can be made from $X_{\phi}$ and $Z_{90}$
(where $U_{\theta}=\cos\frac{\theta}{2}-i \sin\frac{\theta}{2}\,\,U$
and $U^{2}=\id$). 
{\bf Exercise}: show this by showing how to make
$Y_{\phi}$ from $X_{\phi}$ and $Z_{90}$.

The operator $\overline{X}_{\phi}$ is implemented as follows:
\begin{eqnarray*}
\overline{X}_{\phi}=\cos{\frac{\phi}{2}}-i \sin\frac{\phi}{2} \overline{X}
=\cos{\frac{\phi}{2}}-i\sin\frac{\phi}{2} X^{(1)}I^{(2)}
=X^{(1)}_{\phi}I^{(2)}
\end{eqnarray*} 

\begin{figure}[h!]
\begin{center}
\includegraphics[width=0.5\textwidth]{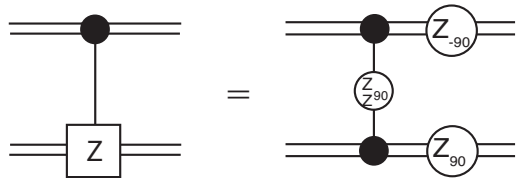}
\end{center}
\caption{\footnotesize The equivalence between the CSign and $(Z^{(1)}Z^{(2)})_{90}$ operation.}
\label{ZZCNOT1Pic}
\end{figure}

The operator $\overline{Z}_{90}$ is implemented as follows:  
\begin{eqnarray*}
\overline{Z}_{90}=\frac{1}{\sqrt{2}}(I-i \overline{Z})
=\frac{1}{\sqrt{2}}(I-iZ^{(1)}Z^{(2)})
=(Z^{(1)}Z^{(2)})_{90}
\end{eqnarray*} 
The $(Z^{(1)}Z^{(2)})_{90}$ operation is closely related to the CSign
as can be seen in fig. \ref{ZZCNOT1Pic}. 
We will show later that teleportation techniques can be used for this gate as it is not error
free.

The $Z_{\pm 90}$ gates can be implemented from error free (in our present model) single qubit rotations.

\subsubsection{Measurements}

After our qubits have been encoded with the code words from eqs.
(\ref{Zmeas0Eqn}) and (\ref{Zmeas1Eqn}), how do we measure in a particular
encoded basis? 

To measure in the encoded $\bar Z$ eigenbasis: measure both qubits in the
$Z$ eigenbasis. The encoded states $\ket{\bar 0}$
and $\ket{\bar 1}$ are distinguished by the parity
of the qubits.  This can be done error free. To measure in the
$\bar Y$ basis we need to rotate the qubit using a $X_{90}$ rotation,
and then measure in the $Z$ basis as mentioned above.
This again can be done error free.  If we wanted to measure in
the $\bar X$ basis, we would need to first rotate the qubits using
a $\bar Y_{90}$.  This rotation, in term of the bare qubits, requires
a $YZ_{90}$ gate and cannot be done error free, it will require a CSign gate
and thus a chance of an error.

\subsubsection{Two qubit rotations}

In order to have a universal set of encoded operations we need an encoded entangling operation.
The $(\overline{Z}\overline{Z})_{90}$ rotation suffices for this. This can be used to make an encoded CNOT, as shown in fig. \ref{ZZCNOT1Pic}. 

\begin{eqnarray*}
(\bar{Z}\bar{Z})_{90}&=&\frac{1}{\sqrt{2}}(1-i\bar{Z}\bar{Z})\\&=&\frac{1}{\sqrt{2}}(1-iZ^{(1)}Z^{(2)}Z^{(3)}Z^{(4)})\\&=&(Z^{(1)}Z^{(2)}Z^{(3)}Z^{(4)})_{90}
\end{eqnarray*}

\begin{figure}[h!]
\begin{center}
\includegraphics[width=0.5\textwidth]{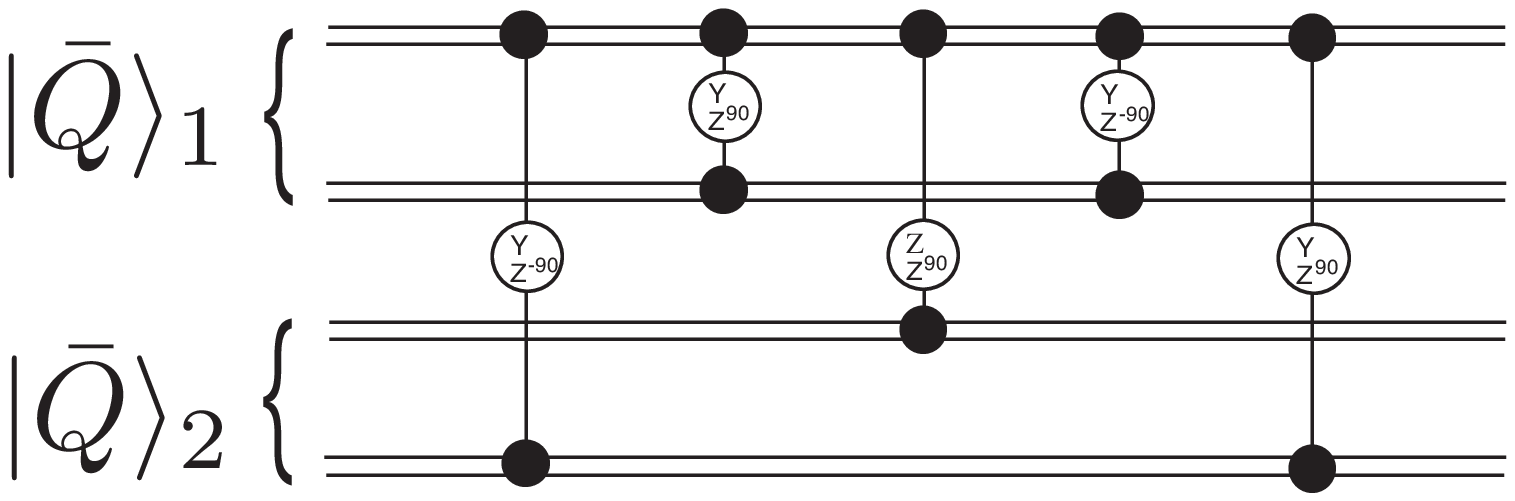}
\end{center}
\caption{\footnotesize A $(\bar{Z}\bar{Z})_{90}$ gate between qubits $\ket{\overline{Q}}_{1}$ and $\ket{\overline{Q}}_{2}$.}
\label{ZZdiag1Pic}
\end{figure}

The circuit to implement $(\overline{Z}\overline{Z})_{90}$ is given in fig. \ref{ZZdiag1Pic}. {\bf Exercise}:  confirm this.

Unfortunately fig. \ref{ZZdiag1Pic} does not readily yield a logical
gate with significantly less error than using the large entangled
states $\ket{t_{n}}$ shown in the last section. To do this requires
teleportation techniques, as will be explained in the following
subsections.

\subsection{Summary so far}

With linear optics we can use teleportation along with the large
entangled states $\ket{t_{n}}$ to perform two qubit operations at the
expense of having $Z$-measurement errors. Without correcting for these
$Z$-measurement errors the probability scales as $\frac{n}{n+1}$.

To correct for these errors we use the encoding
$\ket{\overline{0}}=\frac{1}{\sqrt{2}}\bigl(\ket{00}_{q}+\ket{11}_{q}\bigr)$
and
$\ket{\overline{1}}=\frac{1}{\sqrt{2}}\bigl(\ket{01}_{q}+\ket{10}_{q}\bigr)$. 
Encoded $X$, $Y$ and $Z$ eigenstate preparations are error free as are encoded $Y$ and $Z$ measurements.  

We can perform encoded operations on this code. $\overline{X}_{\phi}$
is error free however $\overline{Z}_{90}$ requires a two qubit gate so
is not. Similarly $(\overline{Z}\overline{Z})_{90}$ is not error free.
 
\subsection{Threshold for $Z$-measurement QECC}

\subsubsection{Accuracy threshold Theorem}

An implementation of information processing is scalable if the
implementation can realise arbitrarily many information units and
operations with fixed accuracy and with physical resource
overheads that are polynomial in the number of information units and
operations.

\textit{Accuracy Threshold Theorem}: If the error per gate is less
than a threshold, then it is possible to efficiently quantum compute
arbitrarily accurately. 

This theorem can be proved by designing fault tolerant gates so that
the error model of the encoded gates is the same as the one of the
bare qubits. Then we need to concatenate these gates. If the error
per qubit at one level of encoding is be less than the error per qubit
at the next level of encoding we can keep this concatenation and
reduce the effective error rate.

\subsubsection{Nice teleportation}

In order to understand how to reach a threshold for quantum error
correction in LOQC we need to first consider the ``nice''
teleportation circuit, shown below in fig. \ref{fig5aKLMPic}. The
reason why this circuit is ``nice'' is because of how it effects
the quantum information when there is an unintentional measurement
error.

\begin{figure}[h!]
\begin{center}
\includegraphics[width=0.6\textwidth]{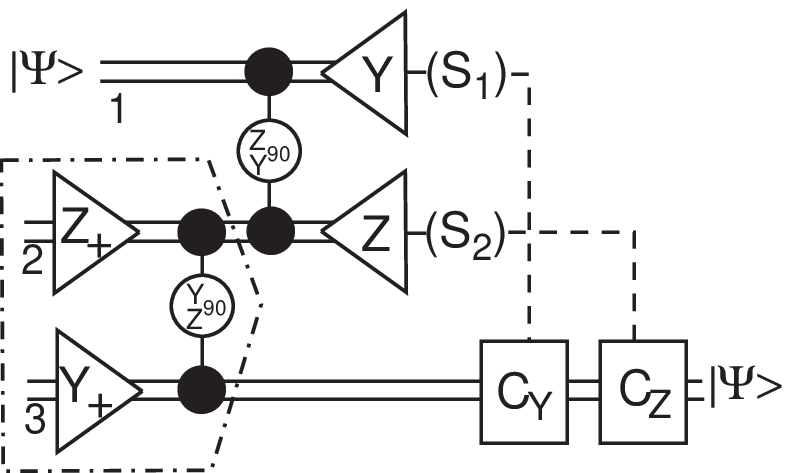}
\end{center}
\caption{\footnotesize A nice teleportation circuit. The state produced in the dotted-dashed box is $\ket{tx_{2}}=\frac{1}{2}\bigl(\ket{00}_{q}+i\ket{01}_{q}+\ket{10}_{q}-i\ket{11}_{q}\bigr)$.}
\label{fig5aKLMPic}
\end{figure}

We want to teleport the state
$\bigl(\alpha\ket{0}_{q}+\beta\ket{1}_{q}\bigr)_{1}$. The action of
the correction gates $C_{Y}$ and $C_{Z}$ for each combination of
$S_{1}$ and $S_{2}$ are given below:
\begin{table}[h!]
\begin{tabular}{cccc}
$S_{1}$&$S_{2}$&$C_{y}$&$C_{z}$\\
\hline 
$-1$&$-1$&$\id$&$\id$\\
$+1$&$-1$&$Z$&$\id$\\
$-1$&$+1$&$\id$&$X$\\
$+1$&$+1$&$Z$&$X$\\
\end{tabular}
\end{table} 
\newpage
{\bf Exercise}:  confirm that if we measure 
\begin{eqnarray*}
\ket{Y^{+}}_{1}\ket{Z^{+}}_{2} &\text{we get}& \bigl(\alpha\ket{1}_{q}-\beta\ket{0}_{q}\bigr)_{3}\\ 
\ket{Y^{+}}_{1}\ket{Z^{-}}_{2} &\text{we get}& \bigl(\alpha\ket{0}_{q}-\beta\ket{1}_{q}\bigr)_{3}\\ 
\ket{Y^{-}}_{1}\ket{Z^{+}}_{2} &\text{we get}& \bigl(\alpha\ket{1}_{q}+\beta\ket{0}_{q}\bigr)_{3}\\ 
\ket{Y^{-}}_{1}\ket{Z^{-}}_{2} &\text{we get}& \bigl(\alpha\ket{0}_{q}+\beta\ket{1}_{q}\bigr)_{3} 
\end{eqnarray*}
before the correction gate. 

If an error occurs during the $Y^{(2)}Z^{(3)}_{90}$ state preparation
gate, we just try again as they do not corrupt the quantum
information. When we consider errors we need to analyse the effect of
an unintended $Z$ measurement error on qubit 1 and the effect of an
unintended $Y$ measurement error on qubit 2 \cite{Knill00}.

\begin{figure}[h!]
\begin{center}
\includegraphics[width=0.6\textwidth]{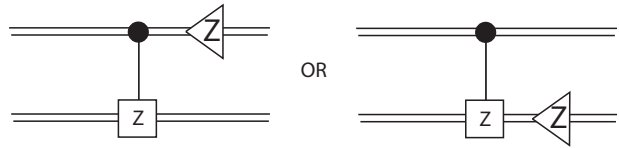}
\end{center}
\caption{\footnotesize Errors that can result from the teleported CSign gate.}
\label{CSignErrPic}
\end{figure}

The motivation behind looking at unintended $Z$ and $Y$ measurement
errors can be seen by looking at the relationship between CSign and
the $ZY_{90}$ gate. First consider a CSign gate. As we saw in section
\ref{BeyondStatePrepSect}, when we use teleportation to implement a
CSign gate it can fail with either a $Z$-measurement error on qubit 1
or a $Z$-measurement error on qubit 2, as seen in
fig. \ref{CSignErrPic}.

\begin{figure}[h!]
\begin{center}
\includegraphics[width=0.6\textwidth]{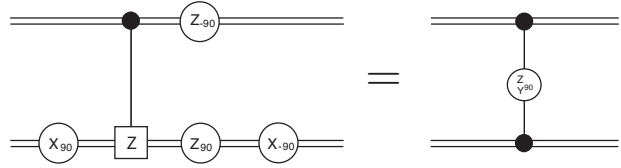}
\end{center}
\caption{\footnotesize Relation between CSign (Control-Z) and $ZY_{90}$.}
\label{ZYCSignPic}
\end{figure}

Now consider how these unintentional $Z$-meansurement errors change
when we look at the $ZY_{90}$ gate. {\bf Exercise}:  using
figs. \ref{CNOTHadPic} and \ref{ZZCNOT1Pic} show that
CSign and $ZY_{90}$ are related as
shown in fig. \ref{ZYCSignPic}. 

Tracing back the potential failures of the $ZY_{90}$ gate due to our
error model using figure~\ref{ZYCSignPic}, we get that they can
originate from a failure of the CSign gate on the first (control)
qubit, leading to a projection in the $Z$ basis or they can
originate from a failure of the CSign gate on the second
(target) qubit, leading to a projection in the $Y$ basis. And, again, this includes the
knowledge of which qubit has failed (as we know this for the CSign
gate). Thus the error model after the teleportation is similar to the
error model for the unencoded qubits.

In more detail, consider the case of a $Z$ measurement error on qubit
1 in fig. \ref{fig5aKLMPic}.

\begin{figure}[h!]
\begin{center}
\includegraphics[width=0.7\textwidth]{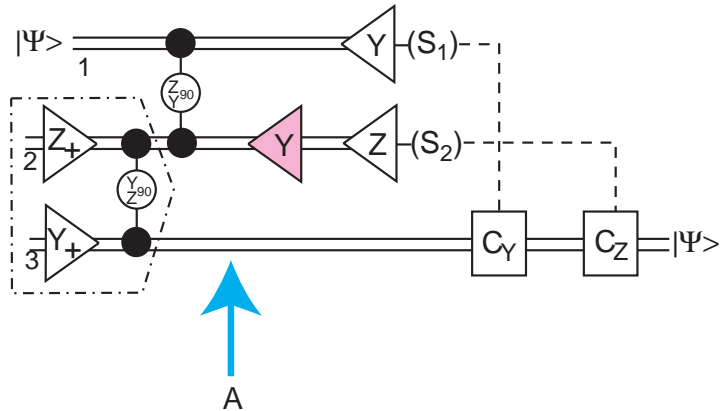}
\end{center}
\caption{\footnotesize Teleportation with a $Y$-measurement error on qubit 2.}
\label{fig5aKLMZ1Pic}
\end{figure}

We need to first work out what the state looks like at point $A$ in 
fig. \ref{fig5aKLMZ1Pic}:  
\begin{eqnarray}
\frac{1}{\sqrt{2}}\Bigl( i\alpha\ket{001}_{q}
+\alpha\ket{010}_{q}+\beta\ket{100}_{q}-i\beta\ket{111}_{q}\Bigr)_{123}
\label{StateatAEqn}
\end{eqnarray}

If we measure qubit 1 in the $Z$ basis we have 
\begin{eqnarray*}
\ket{Z^{+}}_{1}\Bigl(i\alpha\ket{01}_{q}&+&\alpha\ket{10}_{q}\Bigr)_{23}\\
&\text{or}&\\ 
\ket{Z^{-}}_{1}\Bigl(i\beta\ket{00}_{q}&+&\beta\ket{11}_{q}\Bigr)_{23}
\end{eqnarray*}
If this error does occur, our quantum information is corrupted and the
gate failed. The failure mode is again a projection in the $Z$ basis.

Next consider that we have a $Y$ measurement error on qubit 2 in fig. \ref{fig5aKLMPic}. 
Once again the state at point A in fig. \ref{fig5aKLMZ1Pic} is given by 
eq. (\ref{StateatAEqn}). If we measure qubit 2 in the $Y$ basis we have 
\begin{eqnarray*}
\ket{Y^{+}}_{2}\Bigl(i\alpha\ket{01}_{q}-i\alpha\ket{00}_{q}
+\beta\ket{10}_{q}-\beta\ket{11}_{q}\Bigr)_{13}
&=&\ket{Y^{+}}_{2}\Bigl(-i\alpha\ket{0}_{q}
+\beta\ket{1}_{q}\Bigr)_{1}\Bigl(\ket{0}_{q}-\ket{1}_{q}\Bigr)_{3}\\
&\text{or}&\\
\ket{Y^{-}}_{2}\Bigl(i\alpha\ket{01}_{q}+i\alpha\ket{00}_{q}
+\beta\ket{10}_{q}+\beta\ket{11}_{q}\Bigr)_{13}
&=&\ket{Y^{-}}_{2}\Bigl(i\alpha\ket{0}_{q}
+\beta\ket{1}_{q}\Bigr)_{1}\Bigl(\ket{0}_{q}+\ket{1}_{q}\Bigr)_{3}
\end{eqnarray*}
If this error does occur, our quantum information can be recovered and we
can re-use qubit 1 for teleportation.

\subsubsection{Teleportation with error recovery}

Now say we have an encoded qubit
$\alpha\ket{\overline{0}}+\beta\ket{\overline{1}}$ that experiences a
$Z$-measurement error on qubit 1. We saw in fig. \ref{projxx2Pic} that
we could correct for the error. When we teleport each part of the
encoded qubit and then correct for the $Z$-measurement error on qubit
1 we have:

\begin{figure}[h!]
\begin{center}
\includegraphics[width=0.9\textwidth]{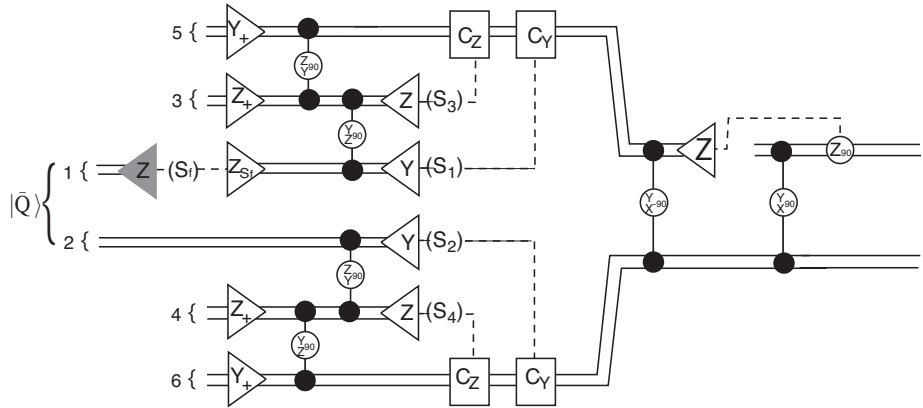}
\end{center}
\caption{\footnotesize Correcting for $Z$-measurement error on qubit 1.}
\label{fig6aKLMPic}
\end{figure}

\begin{figure}[h!]
\begin{center}
\includegraphics[width=0.9\textwidth]{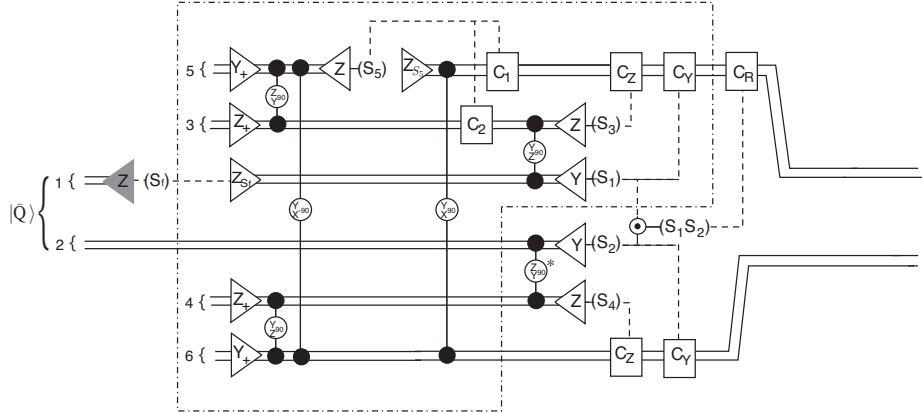}
\end{center}
\caption{\footnotesize Correcting for $Z$-measurement error on qubit 1
  with the error correcting rotations commuted to the state
  preparation stage.}
\label{fig6KLMPic}
\end{figure}

We can commute the $(Y^{(1)}X^{(2)})_{-90}$, measurement and
$(Y^{(1)}X^{(2)})_{90}$ gates to the state production stage, shown in
fig. \ref{fig6KLMPic}.

 {\bf Exercise}: how are the corrections $S_{1}$, $S_{2}$,
$S_{3}$ and $S_{4}$ modified after the $(Y^{(1)}X^{(2)})_{-90}$ and
$(Y^{(1)}X^{(2)})_{90}$ are commuted through?

We take $f$ to be the probability that teleportation fails. We want to
calculate the probability of network \ref{fig6KLMPic} to fail:
$F_{r}$. We can assume the state preparation has succeeded as it is
done offline. We have to look at the modes of failure for the
$(Z^{(2)}Y^{(4)})_{90}$ gate (marked with an asterisk, $*$) after the
state preparation. First we have a term for the probability that qubit
4 will be projected in the $Y$ basis (which fails with a probability
of $f$), which can be undone, and we retry the whole circuit (retrying
fails with a probability of $F_{r}$) and thus we get $f F_{r}$. Next
we have the probability that the $Y$ projection of qubit 4 succeeds
$(1-f)$ but qubit 2 is projected into the $Z$ basis $(f)$ and the
circuit fails: $(1-f)f$. We therefore have
\begin{eqnarray}
F_{r}=f F_{r}+(1-f)f\to F_{r}=f,
\end{eqnarray}
i.e. the probability of this circuit failing is the same
as the probability that the teleportation fails which is the same as the 
corruption of one of the qubits in the CSign gate.
\subsubsection{Encoded $Z_{90}$ Gate}

\begin{figure}[h!]
\begin{center}
\includegraphics[width=0.7\textwidth]{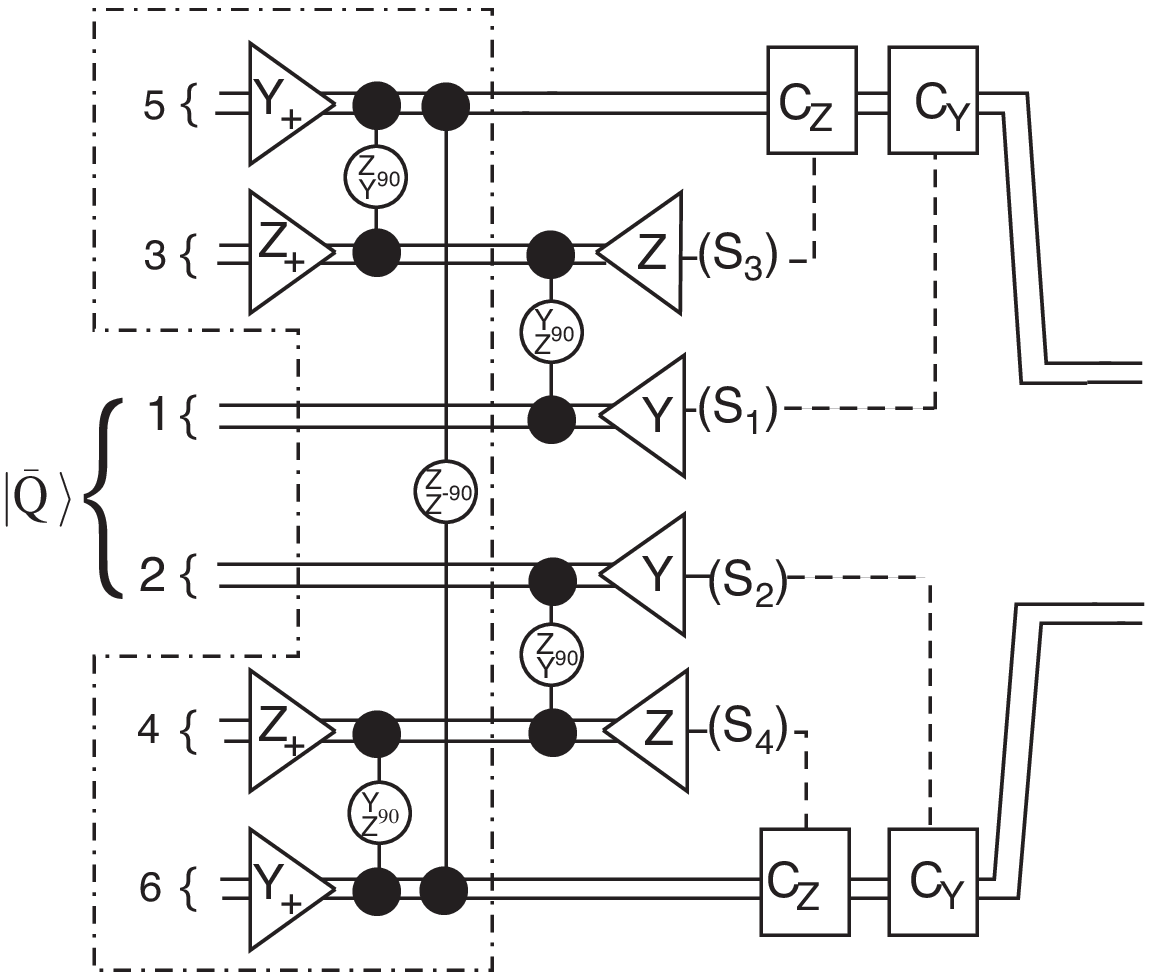}
\end{center}
\caption{\footnotesize An encoded $Z_{90}$ gate:  $\overline{Z}_{90}=(Z^{(1)}Z^{(2)})_{90}$.}
\label{fig5bKLMPic}
\end{figure}

Remember that the encoded $Z_{90}$ gate is given by:
$\overline{Z}_{90}=(Z^{(1)}Z^{(2)})_{90}$. As with the error
correction after teleportation circuit (figs. \ref{fig6aKLMPic} and
\ref{fig6KLMPic}), we can commute the $(Z^{(1)}Z^{(2)})_{90}$ back to
the state preparation stage, as shown in fig. \ref{fig5bKLMPic}.

We want to calculate the total probability of failure of recovery
($F_{z}$) for this circuit. The probability that both teleportations
fail is $f^{2}$. That is, the probability that both qubits 1 and 2 are
projected into the $Z$ basis is $f^{2}$. The probability that one
qubit is successfully teleported but the other is not, and recovery
fails, is $(1-f)f^{2}$. That is, the probability that the top
teleportation succeeds and qubit 4 is projected in the $Y$ basis and
recovery fails with a projection of qubit 2 in the $Z$ basis is
$(1-f)f^2$. The probability that one teleportation succeeds but not
the other, recovery succeeds and we re-attempt, is $(1-f)fF_{z}$. That
is, the probability that the top teleportation succeeds and qubit 4 is
projected in the $Y$ basis and recovery succeeds, so we retry the
entire circuit is $(1-f)fF_{z}$. We therefore have
\begin{eqnarray}
F_{z}=f^{2}+(1-f)f^{2}+f(1-f)F_{z}\to F_{z}=\frac{2 f^{2}-f^{3}}{1-f+f^{2}}
\end{eqnarray}
and when the teleporation fails, the mode of error is a projection in the $Z$
basis.

In order to reduce the probability of failure one would perform the
teleportations sequentially. To perform
$(\bar{Z}\bar{Z})_{90}=(Z^{(1)}Z^{(2)}Z^{(3)}Z^{(4)})_{90}$ we would
simply have the same circuit with two more copies of $\ket{tx_{2}}$. A
similar result for $F_{zz}$ is obtained \cite{Knill00}.

\subsubsection{Threshold}

\begin{figure}[h!]
\begin{center}
\includegraphics[width=0.7\textwidth]{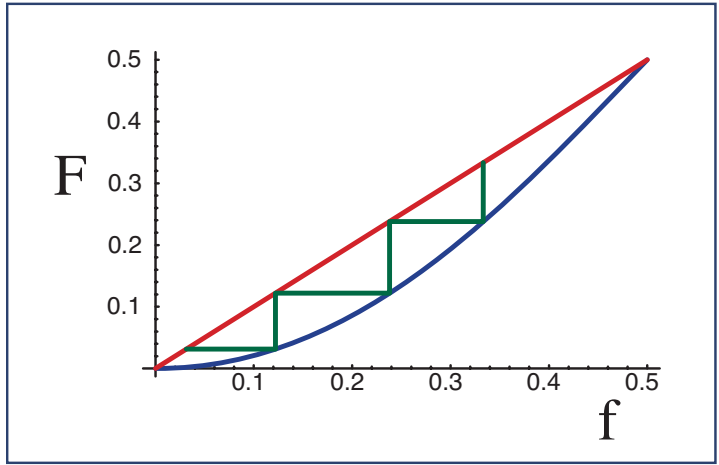}
\end{center}
\caption{\footnotesize Shows $F=f$ and $F=\frac{2 f^{2}-f^{3}}{1-f+f^{2}}$ on the same axis to show the threshold of $f<\frac{1}{2}$.}
\label{threshKLMPic}
\end{figure}

As mentioned earlier, the error per qubit for one level of
concatenation must be less than that for the next level up. In our
case, the error for the $ZZ_{90}$ (or Control-sign gate), $f$, must be
equal to the error for the next level of encoding, $F_{zz}$ (which is
equivalent to $F_{z}$). If we plot $F_{zz}$ vs $f$, we see that
$F_{z}=f$ when $f=\frac{1}{2}$, as can be seen in
fig. \ref{threshKLMPic}. Thus to ensure that 
the fault tolerant error correction procedure reduces 
the error probability we need $f<\frac{1}{2}$.
 
\subsection{Other Errors}

So far we have only considered errors due to incorrect measurements at
the teleportation stage. There are many other types of errors that may
also be important, such as phase errors, inefficient sources, mode
absorption/dispersion, imperfect beam splitters, inefficient
detectors, $\ldots$ We will consider inefficient detector errors in
the next subsection.

\subsubsection{Photon Loss}

\begin{figure}[h!]
\begin{center}
\includegraphics[width=0.7\textwidth]{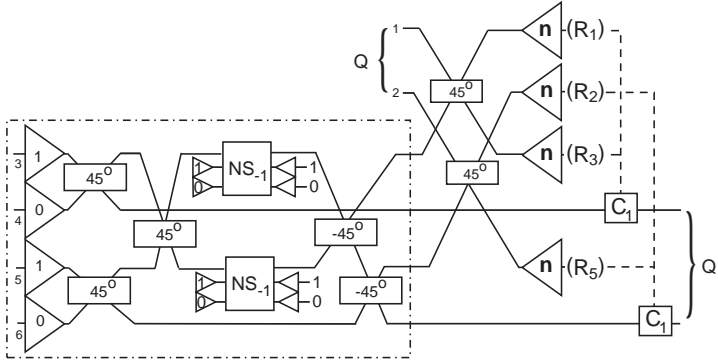}
\end{center}
\caption{\footnotesize  Teleportation with loss detection.}
\label{fig4KLMPic}
\end{figure}

In the original LOQC paper \cite{KLM}, a modification to the gate
teleportation protocol was shown that allowed for the detection of
photon loss, as shown in fig. \ref{fig4KLMPic}.

\begin{figure}[h!]
\begin{center}
\includegraphics[width=0.7\textwidth]{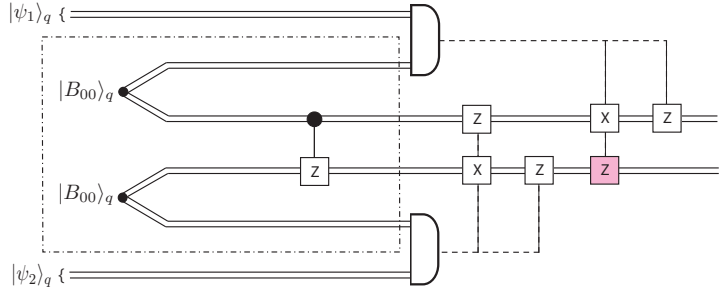}
\end{center}
\caption{\footnotesize A CSign gate with one of the correction gates applied at the incorrect time.}
\label{teleport1Pic}
\end{figure}

Without going through this circuit explicitly, we know where photon
loss occurs through the measurement of ancilla modes. An alternate way
to consider loss in teleportation is to say the qubit being teleported
is fully ``erased'', undergoing the transformation \cite{Silva05}
\begin{eqnarray*}
E(\rho)=\frac{1}{4}\bigl( \rho+X\rho X +Z\rho Z +Y\rho Y \bigr)
\end{eqnarray*}
where $X, Y$ and $Z$ are the usual Pauli operators. This can be
considered to be the depolarising channel conditioned on perfect
information about which qubit is randomised. Consider we have two
state teleportations with the state preparation modified to perform a
CSign gate as in fig. \ref{teleport1Pic}.

We want to consider what happens if we find photon loss on the top
teleporation (for the moment assume there are no $Z$ measurement
errors from the teleportation). The top qubit is fully erased, total
information is lost. But what about the bottom qubit being teleported?

Because the top qubit is completely randomised, when we perform the
gate corrections dependent on the top Bell measurement, the $Z$ gate
(in colour in fig. \ref{teleport1Pic}) will be applied at the
incorrect time. So any photon loss on the top teleportation translates
to a full erasure of that qubit being teleported and a $Z$ erasure of
the bottom qubit:
\begin{eqnarray*}
Z(\rho)=\frac{1}{2}\bigl( \rho+Z\rho Z \bigr)
\end{eqnarray*}

To correct for erasures we can use the 7-qubit code. This has the
stabilzer generators:
\begin{eqnarray*}
\begin{array}{ccccccccccccl}
X&\otimes&X&\otimes&X&\otimes&X&\otimes&\id&\otimes&\id&\otimes&\id\\
X&\otimes&X&\otimes&\id&\otimes&\id&\otimes&X&\otimes&X&\otimes&\id\\
X&\otimes&\id&\otimes&X&\otimes&\id&\otimes&X&\otimes&\id&\otimes&X\\ 
Z&\otimes&Z&\otimes&Z&\otimes&Z&\otimes&\id&\otimes&\id&\otimes&\id\\
Z&\otimes&Z&\otimes&\id&\otimes&\id&\otimes&Z&\otimes&Z&\otimes&\id\\
Z&\otimes&\id&\otimes&Z&\otimes&\id&\otimes&Z&\otimes&\id&\otimes&Z
\end{array}
\end{eqnarray*}
This code can correct for any combination of 1 or 2 erasures and 28 of
the possible 35 combinations of 3 erasures. Some of the combinations of 4 erasures can also be corrected. To recover from an erasure of a qubit in an unknown location we would have to measure all six generators.

\begin{figure}[h!]
\begin{center}
\includegraphics[width=0.4\textwidth]{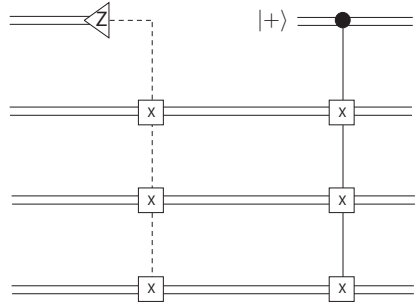}
\end{center}
\caption{\footnotesize A recovery from a $Z$ erasure or $Z$ measurement error on the top qubit. All other qubits must be erasure free in this case. The remaining 3 qubits in the 7 qubit code are not shown.}
\label{zrecoverPic}
\end{figure}

We always know the location of the erasure, reducing the number of
generators to measure down to 2.  The 7-qubit code can also correct
for $Z$ erasures and $Z$-measurement errors. This is because a $Z$
erasure on a particular qubit is equivalent to an unintentional $Z$
measurement on the qubit of unknown outcome: $Z_{+}\rho
Z_{+}+Z_{-}\rho Z_{-}=\frac{1}{2}\bigl(\id+Z\rho Z\bigr)$, where
$Z_{\pm}=\frac{1}{2}\bigl(\id\pm Z\bigr)$. In this case only 1
generator needs to be measured. The circuit for this recovery is shown
in fig. \ref{zrecoverPic} \cite{Silva05}.

In order to estimate how good photon detectors need to be, we estimate the threshold for fault tolerant computation. The gate error threshold was found by Silva \textit{et. al.} \cite{Silva05} to be bounded below by a value between $1.78\%$ and $11.5\%$. $1.78\%$ results from errors due to loss and $11.5\%$ results from $Z$ measurement errors in teleportation.
\newpage
\section{Conclusion}

In these set of lectures we have seen how it was possible to make
linear optical elements with single photon sources and detectors
behave like a quantum computer. A fundamental idea was to use quantum
teleportation to implement gates. If we have probabilistic gates, this
teleportation can transfer the difficulty of making a gate to the
difficulty of making a state. In our case an extra difficulty was that
the teleportation would also induce errors, which could be taken care
of. The whole process was streamlined by using quantum error
correction, the subject of the last section.

The work on linear optics raises interesting questions about what quantum information is and the origin of its power. Linear optical elements by themselves can be simulated efficiently on a classical computer, but when we measure and condition further gates constructed from linear elements, this is not possible anymore. It is somewhat surprising that this condition on classical information becomes so powerful.

In these lectures we have not mentioned the experiments that have demonstrated the first steps towards  implementing LOQC. This is a very active and interesting subject that will be left to another set of lectures.

\newpage

\acknowledgments We thank E. Knill, G. Milburn, M. Ericson, K. Banaszek and M. Silva for useful
interaction. CRM and RL are supported in part by CIAR, NSERC, and CIPI, RL is
also supported in part by ORDCF and the Canada Research Chair program.

\bibliographystyle{unsrt}

\end{document}